\documentclass[conference]{IEEEtran}
\usepackage[T1]{fontenc}

\usepackage{cite}
\usepackage{amsmath,amssymb,amsfonts}
\usepackage{graphicx}
\usepackage{textcomp}
\usepackage{xcolor}
\def\BibTeX{{\rm B\kern-.05em{\sc i\kern-.025em b}\kern-.08em
    T\kern-.1667em\lower.7ex\hbox{E}\kern-.125emX}}
    
\usepackage[utf8]{inputenc}
\usepackage{graphicx}
\usepackage{paralist}
\usepackage{subcaption}
\usepackage{algorithm}
\usepackage[noend]{algpseudocode}
\usepackage{multirow}
\usepackage{adjustbox}
\usepackage{acronym}
\usepackage{amsmath}
\usepackage{longtable}
\usepackage{xurl}
\usepackage{booktabs}
\usepackage{tikz}
\newcommand*\circled[1]{\tikz[baseline=(char.base)]{
            \node[shape=circle,draw,inner sep=1pt] (char) {#1};}}

\usepackage{array}
\newcolumntype{P}[1]{>{\centering\arraybackslash}p{#1}}

\usepackage{ulem}
\usepackage{color,colortbl}
\usepackage{lineno}
\newcommand\redsout{\bgroup\markoverwith{\textcolor{red}{\rule[0.5ex]{2pt}{0.4pt}}}\ULon}

\newif\ifcomments
\commentstrue
\begin{document}

\title{Sniper Backdoor: Single Client Targeted Backdoor Attack in Federated Learning}

% \thanks{The European commission financially supported this work through the Horizon Europe program under the IDUNN project (grant agreement number 101021911). It was also partially supported by the Ayudas Cervera para Centros Tecnológicos grant of the Spanish Centre for the Development of Industrial Technology (CDTI) under the project EGIDA (CER-20191012) and by the Basque Country Government under the ELKARTEK program, project TRUSTIND - Creating Trust in the Industrial Digital Transformation (KK2020/00054).
% }

\author{\IEEEauthorblockN{1\textsuperscript{st} Gorka Abad}
\IEEEauthorblockA{\textit{Radboud University, Nijmegen, The Netherlands} \\
\textit{Ikerlan Technology Research Centre, Arrasate-Mondragón, Spain}\\
0000-0002-6735-3623}
}
% \author{
%         Gorka Abad\inst{1,2}\orcidID{0000-0002-6735-3623} \and
%         Servio Paguada\inst{1,2}\orcidID{0000-0003-4665-7457} \and
%         Stjepan Picek\inst{1}\orcidID{0000-0001-7509-4337} \and
%         Víctor Julio Ramírez-Durán\inst{2}\orcidID{0000-0002-7384-9908} \and
%         Aitor Urbieta\inst{2}\orcidID{0000-0001-5836-4198}
%         }

% \authorrunning{G. Abad et al.}

% \institute{Radboud University, Nijmegen, The Netherlands \and
%             Ikerlan Technology Research Centre, Arrasate-Mondragón, Spain}

\maketitle

\begin{abstract}

Federated Learning (FL) enables collaborative training of Deep Learning (DL) models where the data is retained locally. Like DL, FL has severe security weaknesses that the attackers can exploit, e.g., model inversion and backdoor attacks. Model inversion attacks reconstruct the data from the training datasets, whereas backdoors misclassify only classes containing specific properties, e.g., a pixel pattern. Backdoors are prominent in FL and aim to poison every client model, while model inversion attacks can target even a single client.

This paper introduces a novel technique to allow backdoor attacks to be client-targeted, compromising a single client while the rest remain unaltered. The attack takes advantage of state-of-the-art model inversion and backdoor attacks. We leverage a Generative Adversarial Network to perform a model inversion attack. Afterward, we shadow-train the FL network, in which, using a Siamese Neural Network, we can identify, target, and backdoor the victim's model.
Our attack has been validated using the MNIST, F-MNIST, and EMNIST datasets under different settings---achieving up to 99\% accuracy on both source (clean) and target (backdoor) classes and against state-of-the-art defenses, e.g., Neural Cleanse, opening a novel threat model to be considered in the future.

% \keywords{Federated Learning \and Backdoor attacks \and Model inversion \and Client-wise attack}

\end{abstract}

\section{Introduction}
\label{sec:introduction}

Deep learning (DL) achieves state-of-the-art performance in various machine learning tasks, e.g., computer vision~\cite{wu2019wider}, speech recognition~\cite{greff2016lstm}, and natural language processing~\cite{brown2020language}. Unfortunately, DL has severe security and privacy flaws, which have been exploited in recent years, e.g., backdoors~\cite{gu2019badnets} and inference attacks~\cite{nasr2019comprehensive}. Therefore, research has also focused on creating secure DL algorithms that prevent leaking data or causing misbehavior~\cite{gao2020backdoor}. 

Contrary to centralized DL algorithms, where data and the model are stored in a single point and trained, in 2016, Google developed a privacy-preserving decentralized and collaborative training approach, i.e., \textit{Federated Learning} (FL)~\cite{mcmahan2017communication}. FL is composed of a set of clients and an aggregator (server), where clients store their data and train the same DL model locally for a couple of epochs. Then, the model parameters are shared with the aggregator, which merges them, joining the properties of heterogeneous datasets without accessing them.

FL protocols, though their decentralized structures, are also prone to security and privacy attacks~\cite{abad2021on,mothukuri2021survey}. 
Among the most popular attacks in FL are \textit{backdoor} and \textit{inference} attacks. While backdoor attacks focus on modifying the training set to cause behavior at inference time~\cite{gu2019badnets, kairouz2019advances}, inference attacks extract private information from the DL model~\cite{chen2020beyond,song2020analyzing}. 
More precisely, backdoor attacks focus on altering the training set by including specific triggers in the input space, which cause misbehavior of the DL model only under the presence of the trigger. If it is not present, the DL model will behave normally. Backdoor attacks have been adapted to FL, where modifying a single client dataset can cause the joined model to misbehave, poisoning each client in the network~\cite{Bagdasaryan2020, sun2019can, wang2020attack}.

Note that the existing backdoor attacks on FL cause all clients to end up with a backdoored model. 
However, there could be scenarios where the server targets a single client or a subset of the clients, not all.
%In other words, the server aims to poison only the targeted (victim) clients.
For example, imagine a scenario where competitive banks want to jointly train a model for credit scores.
Among those banks, one is targeted by the attacker who may support or remain neutral to the other banks and thus, does not want to backdoor them. 
This unexplored single target scenario gives rise to the following question: 
\textit{Is it possible to launch a backdoor attack, where only targeted (victim) clients get a backdoored model whereas the remaining (non-victim) clients get a clean model?}

%However, this procedure is noisy and unsuitable for FL setups where partners and victims form the network, which causes partners also to get a poisoned model from the aggregator. 

%Motivated by this gap in the ability of the backdoor to adapt to this situation, we investigated a backdoor attack that can be client-targeted, only poisoning a subset of clients, while the models of the partners remain unaltered. The attack combines state-of-the-art inference and backdoor attacks to recover datasets from clients, identify the target clients, and inject the backdoor.

\subsection{Our Contributions}

In this work, we positively answer this question by providing a backdoor attack that can be client-targeted, only poisoning a subset of clients while the models of the other clients remain unaltered.
Our attack combines state-of-the-art inference and backdoor attacks to recover datasets from clients, identify the target clients, and inject the backdoor. Precisely, to achieve a targeted backdoor, we redesign the existing inference attacks by making them client-specific, allowing the generation of client-like datasets. Furthermore, we create a shadow network in FL, which in combination with an SNN, allows identifying the victim among all the clients.
This results in a client-targeted backdoor, where the victim gets a poisoned model while the rest of the clients get a clean one.
%To the best of our knowledge, research has not yet contemplated the viability of backdoors attacks for a target client without compromising the rest. This study examines the possibility of client-wise targeted backdoors as discussed in the threat model. 
Our main contributions are:
\begin{compactitem}
    \item We present, to the best of our knowledge, the first client-targeted backdoor attacks for FL settings, which first identifies the victim, then injects the backdoor only for the victim, but not the other clients. 
    \item We adapt and train a Siamese Neural Network (SNN) for their usage with triplet-loss, which enables identifying anonymous clients in the FL network with up to 92\% accuracy.
    \item We extend and analyze backdoor attacks' capabilities, injecting the trigger at near convergence and focusing on a target client, achieving up to 99\% accuracy on the clean and backdoor test sets and confirming the viability of our attack.
    \item We apply state-of-the-art defense mechanisms to our protocol. Since most of the defenses are on the server side or do not fit our setup, we modified them to fit our attack scenario, e.g., we adapted Trojanzoo, a backdoor testing framework, to utilize our attack. More precisely, we consider Neural Cleanse and smoothing.
\end{compactitem}

To improve the reproducibility, we share the source code of our work.\footnote{\url{https://anonymous.4open.science/r/Sniper-Backdoor-42BD}}

\subsection{Related Work}

%\onote{The explanation of what is backdoor can be removed. Here, we should explain how the existing backdoor attacks applied (very brief), what are their differences, backdoors on FL, and finally defenses}

FL has gained attention as a privacy-driven alternative to centralized learning---granting the ability to train a DL algorithm without sharing the data and splitting the computational power~\cite{singh2019detailed}. %FL further leverages this idea to make DL privacy-driven. 
However, it has been shown that FL is also vulnerable to attacks that make the DL models misbehave at inference time, i.e., poisoning or backdoor attacks~\cite{Bagdasaryan2020, shafahi2018poison,gu2019badnets}, or causing privacy leakages, i.e., inference attacks~\cite{nasr2019comprehensive}. 
Comprehensive studies about the attacks on FL are given in~\cite{abad2021on,mothukuri2021survey}.
In this work, we focus on inference and backdoor attacks.

%Inference attacks' goals vary depending on the attacker's capabilities, e.g., oracle access to the model, access to the inner computations, or the model exchanges between clients and the aggregator. Nasr et al.~\cite{nasr2019comprehensive} developed an attack that gets if a data sample has been used during training, i.e., membership inference. Further, FL has also utilized membership attacks, indicating if a specific client has used a concrete data sample~\cite{chen2020beyond,song2020analyzing}. Other attacks, such as model inversion, reconstruct the training set of a DL model by inspecting its weights~\cite{fredrikson2015model}. Furthermore, Song et al. leveraged a Generative Adversarial Network (GAN) to create images similar to the ones used during the training for a specific client~\cite{song2020analyzing}.

%A backdoor is a particular type of poisoning attack whose goal is to misclassify a sample just under a presence of a property or a characteristic, e.g., a single modified pixel or a pixel pattern~\cite{gu2019badnets}. In the computer vision domain, backdoors alter the training set of a single (or a subset) client or directly modify the model. Attacks are classified as targeted or untargeted if the attacker will target a specific label or create uncontrolled misclassification~\cite{mothukuri2021survey}. 

Inference attacks target to learn private information about the training process, which can be either to check if a data sample is used in the training set (membership inference)~\cite{nasr2019comprehensive,chen2020beyond,song2020analyzing} or to reconstruct the training set (model inversion)~\cite{fredrikson2015model}.
%Recently, there have been inference attacks applied to FL setting as well~\cite{chen2020beyond,song2020analyzing}.
A backdoor is a particular type of poisoning attack whose goal is to misclassify a sample just under a presence of a property or a characteristic.
Backdoor attacks have been widely mitigated for different machine learning tasks, e.g., audio or images~\cite{wang2019neural, udeshi2022model, chen2018detecting, liu2019abs}.
In the FL learning setting, the backdoor attack can be applied by the clients  %\todo{but don't we say that no one did this up to our paper?} \onote{this is for the attacker, the attacker can be some of the clients, but the victim was all of them. Anyway, I removed "(some of)" } 
or the aggregator~\cite{Bagdasaryan2020, sun2019can, wang2020attack,xie2019dba}.
The backdoor attack in the FL setting is more challenging than in the centralized setting since each client has a weighted effect on the global model.
To overcome this issue, attackers can either amplify the weights of backdoored model~\cite{Bagdasaryan2020} or distribute the backdoor into several clients~\cite{xie2019dba}.
The existing backdoor attacks on FL settings degrade the global model for all clients, whereas, to the best of our knowledge, there is no client-targeted backdoor attack.

In addition to the attacks, there have been several defense protocols proposed to mitigate them. 
In general, the defenses are based on either dataset inspection~\cite{chen2018detecting,udeshi2022model} or model inspection~\cite{wang2019neural,liu2019abs}. 
Moreover, there are defense mechanisms for FL setting~\cite{blanchard2017machine,fung2018mitigating,andreina2021baffle,xie2021CRLF,fu2019attack}, which are mostly relying on a noise addition or alternative reweighting done by an honest aggregator.
%Despite developing defense mechanisms, many security and privacy issues still exist~\cite{abad2021on}.
The details of the defense mechanisms are given in Section~\ref{sec:defense}.

% During the training phase, backdoor attacks inject poisoned samples into the training dataset or alter the model, directly downgrading every model's performance in the FL network~\cite{Bagdasaryan2020}. Whether the attack's goal is to downgrade the model's overall accuracy or just some target class, the attack is classified as untargeted or targeted~\cite{mothukuri2021survey}. Researchers investigated backdoors in FL scenarios where they scale up the model weights to bypass the vanishing effect caused by the FL aggregation~\cite{Bagdasaryan2020}. Similarly, Xie et al. distributed the backdoor into some clients, where every piece of the backdoor merges after the aggregation, creating a fully backdoored model~\cite{xie2019dba}.

\section{Background}
\label{sec:background}

This section starts with an overview of deep learning and federated learning. Afterward, we discuss backdoor attacks, inference attacks, Generative Adversarial Networks, and Siamese Neural Networks.

\subsection{Deep Learning \& Federated Learning}

\paragraph{Deep Learning} DL algorithms are parameterized functions $\mathbb{F}_\theta$ that maps an input $\textbf{x}\in \mathbb{R}^a$ to an output $y \in \mathbb{R}^b$. 
In the image domain, the dataset is constructed from a collection of images and their labels $\{\textbf{x}, y\}^n$ of size $n$ where $\textbf{x}$ is a vector of pixel values, and $y$ is a vector of probabilities of being from a class $c \in C$.
The parameters, $\theta$, are iteratively set by finding the optimal value for which $\mathbb{F}_\theta(\textbf{x}) = y$, achieved by training. 
During training, large sets of data are provided, and the distance from the predicted output $\mathbb{F}_\theta(\textbf{x)}$ to the ground truth value $y$ is measured---penalizing predictions that are far away leveraging a loss function $\mathcal{L}$. 
Therefore, the optimal values of the parameters $\theta'$ are given by the following equation:
\begin{equation}
\label{eq:training}
      \theta' = \underset{\theta}{\mathrm{argmin}} \sum_{i=1}^n \mathcal{L} (\mathbb{F}_\theta (\{\textbf{x}_i, y_i \})).  
\end{equation}

\paragraph{Federated Learning} FL is a privacy-driven decentralized scheme for collaborative training of ML models. It was introduced by Google, where they proposed creating a network of clients who own distinct datasets to train a global model without directly sharing their datasets~\cite{mcmahan2017communication}. The network is composed of a server (aggregator) and $N$ clients. Every participant of the network, upon consensus, decides to train the same model $W$ under the same conditions, e.g., learning rate (\textit{LR}) and the number of epochs. After local training, following Eq.~\eqref{eq:training}, clients upload their model updates $u_t$ (the differences with the previous epoch) to the aggregator, who joins them by averaging, and sends the new model $W_{t+1}$ back to each client. The FL procedure is repeated for $t$ epochs until convergence is reached:
\begin{eqnarray*}\label{eq:fl}
    W_{t+1} \leftarrow W_t + \frac{1}{N} \sum_{i=1}^{N} u_{t+1}^i.
\end{eqnarray*}

\subsection{Backdoor Attacks}

Backdoor attacks compromise DL networks during training, causing misbehavior at inference time by injecting poisoned samples into the training set. 
A poisoned sample has a trigger embedded over the clean data sample that assigns a target label $\hat{y} \in C$ different from the ground truth label $\hat{y} \neq y$. The attacks are classified as targeted or untargeted if the attacker will target a specific label or create uncontrolled misclassification~\cite{mothukuri2021survey}.

In the image domain, the backdoor trigger is usually a modified pixel or group of pixels, e.g., a pixel square, with a given size and position, e.g., left-top corner or center. The percentage of poisoned samples in the training set is controlled by $\epsilon = \frac{m}{n+m}$ where $m\ll n$ and the poisoned set is shown by $\hat{D}_{train}$. Here, a small $\epsilon$ value usually implies that it is more challenging to include the backdoor behavior, but it causes a stealthier attack. 
During the training procedure with $\hat{D}_{train}$, the backdoor effect is injected into the DL algorithm given by the loss function taking into account the backdoor accuracy:
\begin{equation*}
\label{eq:backdoor}
      \theta' = \underset{\theta}{\mathrm{argmin}} \sum_{i=1}^n \mathcal{L} (\mathbb{F}_\theta (\{\textbf{x}_i, y_i \})) + \sum_{j=1}^m \mathcal{L} (\mathbb{F}_\theta (\{\hat{\textbf{x}}_j, \hat{y}_j \})).  
\end{equation*}

In FL, when a backdoored model is submitted to the server, the poison weights vanish due to aggregation since a single vector of outline values gets averaged with the rest of the clients, thus making the poisoned weights less relevant. However, the vanishing effect can be overcome by upscaling the weights of the model~\cite{Bagdasaryan2020, sun2019can, wang2020attack}.

\subsection{Inference \& Model Inversion Attacks}

Inference attacks measure information leakage through a DL model about its training data. There are different ways in which the attacker obtains private information; for example, by observing the input and outcome of a DL model or observing the inner computation if the attacker has access to the model~\cite{nasr2019comprehensive}. In FL scenarios, the attacker could be either the aggregator who has access to individual updates over the epochs or the clients who have access to the joined model and can control the parameters of the model to update. Depending on the attacker's knowledge and capabilities, the attacker can perform a passive attack only by observing the computations or an active attack that modifies the model's parameters. Similarly, model inversion attacks exploit confidence values obtained during the predictions for reconstructing data from the training dataset of a DL model~\cite{fredrikson2015model}. Furthermore, in FL, generative adversarial networks (GANs) enable the recovery of user-specific data records~\cite{song2020analyzing, chen2020beyond}.

\subsection{Generative Adversarial Networks (GANs)}
\label{sec:GAN}

Deep learning aims to create rich models representing the probability distribution of different data. In contrast, deep generative models aim to generate data samples with a similar distribution as those provided. So far, deep generative models leverage the max-likelihood estimation for such a task. 
However, approximating the probability computation on the max-likelihood estimation is hard. Goodfellow et al. developed a framework for estimating generative models via an adversarial process~\cite{goodfellow2014generative}:
\begin{equation*}
    \label{eq:maxlikelihood}
    J^{(G)} = -\frac{1}{2}\mathbb{E}_z\mathrm{exp}(\sigma^{-1}(D(G(\textbf{z})))),
\end{equation*}
where $J$ is the cost function, $\sigma$ is the logistic sigmoid function, $G$ is the generator, and $D$ is the discriminator. $G$ and $D$ are the two parties involved in generative adversarial networks (GANs), which follow an estimation process based on simulation training via a \textit{zero-sum game}, also called \textit{minimax} game:
\begin{equation*}
    \label{eq:minimax}
    \theta^{(G)'} = \underset{\theta^{(G)}}{\mathrm{argmin}} \underset{\theta^{(D)}}{\mathrm{max}}V(\theta^{(D)},\theta^{(G)}).
\end{equation*}

$G$ takes noise $\textbf{z} \sim p_Z$ samples from some distribution and creates actual data samples, while $D$ distinguishes fake samples from real ones. Both train simultaneously until achieving the Nash equilibrium, where $G$ can generate real-enough data samples that $D$ cannot differentiate. Thus, the distribution of the generated fake samples $p_{G(Z)}$ converges towards the distribution of real data samples. 
GANs have been widely used in different domains, e.g., image creation~\cite{ramesh2021zero}, NLP~\cite{brown2020language}. Concerning security, GANs have also achieved an important role, performing inference attacks~\cite{zhang2020gan} or generating adversarial examples~\cite{xiao2018generating}.

\subsection{Siamese Neural Networks (SNNs)}

SNN is a type of architecture constructed by two identical networks (having the same parameters and structure) with the aim of finding similarities in inputs by comparing their feature vectors' latent space~\cite{bromley1993signature}. 
Since SNN involves pairwise data for training, the loss function has to optimize the model to minimize the distance, e.g., Euclidean distance, between similar inputs and maximize it between different inputs. Triplet loss~\cite{schroff2015facenet} improves learning embeddings of inputs, and it is usually used to improve SNN performance during training.
The same class inputs are close together (in the embedding space) while different class inputs are well separated.\footnote{Similar to Support Vector Machines~\cite{cortes1995support}, where some margin separates different classes' samples.} 
Triplet networks gained popularity with the development of FaceNet~\cite{schroff2015facenet}. Since then, triplet networks have been used in diverse domains, e.g., side-channel analysis~\cite{wu2021best} or learning image similarity~\cite{wang2014learning}.

Triplet loss ($\mathcal{L}$) is formally defined as a function maximizing the distance of the embedding space regarding the anchor ($A$), positive ($P$), and negative ($N$) samples by some margin $\alpha$. Since $A$ and $P$ samples have the same label, triplet loss optimizes the model so that the distance between $A$ and $N$ samples is more significant than between $A$ and $P$:
\begin{equation*}
    \mathcal{L}_{A,P,N} = \max(||f(A) - f(P)||^2 - ||f(A) - f(N)||^2 + \alpha, 0).
\end{equation*}
During the $\mathcal{L}$ minimization, $||f(A) - f(P)||^2$ is pushed towards $0$, while $||f(A) - f(N)||^2$ is being larger than $||f(A) - f(P)||^2 + \alpha$.
The loss can also be represented in different forms according to the three triplet categories: (i) \textit{Easy triplets} where the negative sample is sufficiently distant from the anchor compared to the positive sample to the anchor, (ii) \textit{Hard triplets} where the distance between the negative sample and the anchor is smaller than the positive to the anchor, and (iii) \textit{Semi-Hard triplets} where the distance between the negative and the anchor is larger than the positive to the anchor but is not larger than a margin~$\alpha$, i.e., $||f(A) - f(P)||^2<||f(A) - f(N)||^2 < ||f(A) - f(P)||^2 + \alpha$.

Triplet samples are constructed via online triplet mining, where the anchor, positive, and negative samples are computed on the fly for every batch of inputs. A correct selection of triplets will influence the quality of the model. Thus, in our research and following the suggestions of the original paper~\cite{schroff2015facenet}, we use the \textit{semi-hard} triplets.

\section{Threat Model}

This section discusses the scenario we follow and the assumptions we make. Afterward, we provide details about the adversary's objectives and capabilities.

\subsection{Our Scenario and Assumptions}

In the FL setting, we have clients who train their local DL models with their private datasets and the aggregator (server) who combines the local DL models into a global one and updates the clients.
In our threat model, the attacker is placed on the server, and the victim is among the clients.
It is important to stress that only one client is the targeted victim.
The victim, like any other client, aims to obtain a trained global DL model that has been fed by all clients' datasets.
The attacker aims to poison the joined model to inject a backdoor into the victim's DL model and yet deliver a clean DL model for the other clients.

% In our scenario, we consider a three parties set up: the \textit{victim}, the \textit{clients}, and the \textit{attacker}. The victim is the dataset owner who wants to train a DL algorithm. The victim and the clients train a DL algorithm following the FL process \onote{Victim is also client right? maybe it should be cleared}. The attacker is placed on the server and can access the victim's and clients' provided models and the aggregation procedure\onote{Can the attacker access to their model? or only the aggregation data?} \gnote{To the model they submit, not their dataset.}. The attacker aims to poison the joined model to inject a backdoor into the victim's DL model. 

In FL model training, unlike centralized ones, clients usually have different datasets, i.e., Non-IID (\textit{independent and identically distributed}), where each client has disjoint labeled data samples, i,e., horizontal FL.
%Centralized ML joins data from different sources into a single point for training DL algorithms, i.e., \textit{independent and identically distributed data} (IID). However, in FL, clients have datasets that do not overlap, i.e., non-IID, where each client has disjoint labeled data samples. 
However, to be comprehensive, we consider both IID, i.e., vertical FL and Non-IID dataset cases, which remain private for each client. 
We assume that clients simultaneously share their models anonymously with the aggregator, i.e., their identity cannot be matched to the model submitted. For example, clients could leverage Tor~\cite{dingledine2004tor} for model anonymization. Note that recent works also followed these assumptions~\cite{chen2020beyond,song2020analyzing}.

\subsection{Adversarial Objectives \& Capabilities} 
\label{sec:advobjectives}

We consider a set of clients who wishes to train a DL algorithm for finding the optimal parameters, $\mathbb{F}_{\theta'}$, each of them using a non-colluding dataset $D_{train}$. Each client shares $\mathbb{F}_\theta$ with the aggregator, which returns the joined parameters $\theta_{join}$ after each epoch. 
Before achieving convergence of the joined model, the attacker injects and returns the victim a set of malicious parameters that include the backdoor behavior $\hat{\theta} \leftarrow \theta'$. The attacker has access to the training function, $F_{\theta}$, and has to carefully tune the training process to determine the best values for $\hat{\theta}$. The attacker poisons the holdout training dataset $\hat{D}_{train}$ including triggers in the samples, and obtains the poisoned parameters $\hat{\theta} \leftarrow \mathbb{F}_\theta(\hat{D}_{train})$. For a successful attack, the attacker should achieve high accuracy on the main task, i.e., on the client's $D_{val}$ and the backdoor task.

To evaluate the performance of the attack, we utilize three metrics: 
\begin{compactenum}
    \item \textbf{Clean accuracy:} measures the overall accuracy of the backdoor model over a clean test set.
    \item \textbf{Source class accuracy:} measures the source class accuracy over a clean test set.
    \item \textbf{Target class accuracy:} measures the target class accuracy over a fully poisoned test set, also known as Attack Success Rate (ASR).
\end{compactenum}

\section{Proposed Client-wise Targeted Backdoor}

In this section, we start with an overview of the proposed attack. Afterward, we discuss each attack's component.

\subsection{Attack Overview}

\begin{figure*}[!htb]
    \centering
    \includegraphics[width=\linewidth]{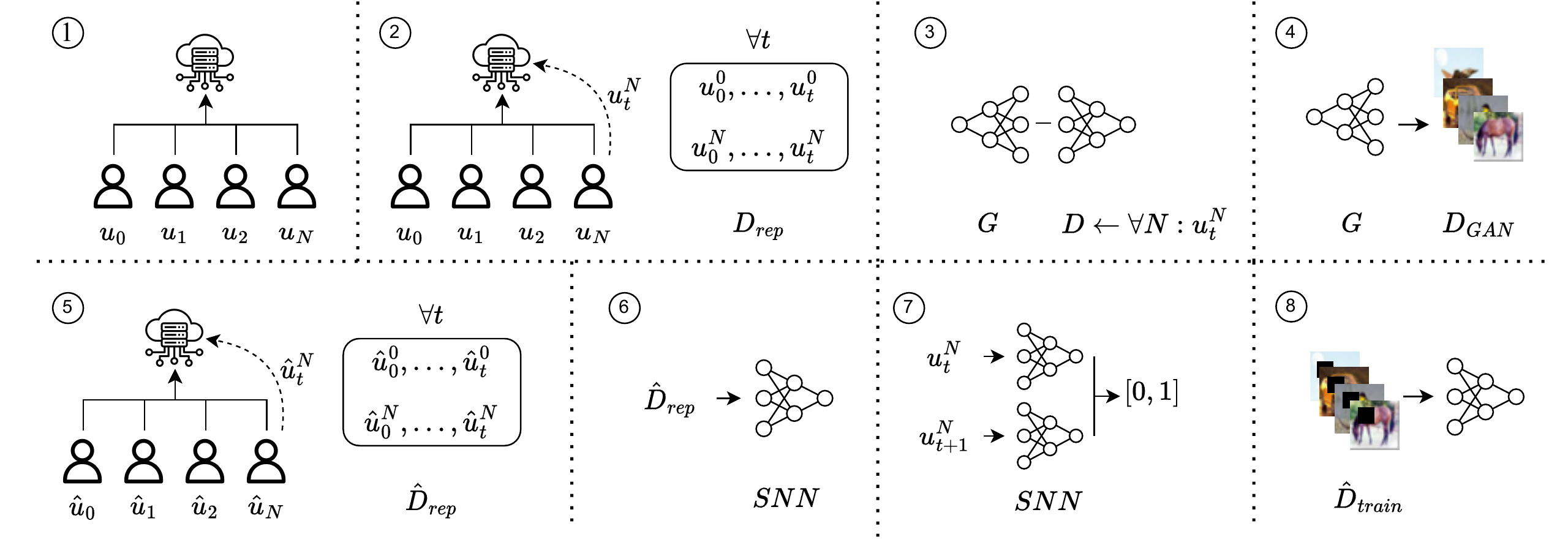}
    \caption{Attack overview.}
    \label{fig:attack}
\end{figure*}

Our attack aims to (i) identify the victim from the received anonymous updates via inference attacks, and then (ii) perform a backdoor attack on the victim.
%As an overview of the attack (see Figure~\ref{fig:attack}), the aim is to identify the anonymous updates via inference attacks based on~\cite{chen2020beyond,song2020analyzing} and gather dataset samples from each client to perform the backdoor attack later. 
As seen in Figure~\ref{fig:attack}, the attack begins with a standard FL training procedure depicted in step \circled{1} where the attacker saves anonymous clients' updates at each FL epoch in step \circled{2} (Section~\ref{sec:train}).
The attacker then uses each client's model at epoch $t$ for constructing a GAN shown in step \circled{3}. The influence of choosing an appropriate $t$ is discussed in Section~\ref{sec:creatingData}.
% Once convergence is met, the attacker selects a set of clients' models at epoch $t$, which influence we discuss in Section~\ref{sec:GAN}.
In step \circled{4}, each client's synthetic dataset is generated using a GAN. To identify anonymous clients, we create a replica of the FL network (Section~\ref{sec:train}), namely the shadow network (Section~\ref{sec:shadow}), which gives the attacker the full details of the training procedure. In step \circled{5}, the attacker trains the shadow network and, in step \circled{6}, records identified updates, which are used to train an SNN in step \circled{7}. The SNN (Section~\ref{sec:SNN}) is used to identify anonymous updates recorded during the original FL network training. In step \circled{8}, the victim client is identified, and the information needed is now acquired to create the backdoor model in step \circled{9} and send it to the victim (Section~\ref{sec:backdoor}).

% As an overview of the inference phases (see Figure~\ref{fig:attack}), the aim is to identify the anonymous updates via inference attacks based on~\cite{chen2020beyond,song2020analyzing} and gather dataset samples from each client to perform the backdoor attack later. (1) The attack begins with a standard FL training procedure, (2) where the attacker saves anonymous clients' updates at each FL epoch (Section~\ref{sec:train}).
% Once convergence is met, the attacker selects a set of clients' models at epoch $t$, which influence we discuss in Section~\ref{sec:GAN}.\todo{how he selects them?} 
% For each model, (3) the attacker constructs a GAN where the discriminator is the client model at epoch $t$ (Section~\ref{sec:GAN}). Thus, one of the necessary information pieces has been fulfilled (4) by creating each client's synthetic dataset using a GAN. A structurally identical \todo{identical to what?} shadow FL network is constructed to identify updates (Section~\ref{sec:shadow}). (5) The attacker trains the shadow network and (6) records identified updates, (7) which are used to train an SNN (Section~\ref{sec:SNN}). The SNN is then used to identify anonymous updates recorded during the original FL network training (Section~\ref{sec:identification}). (8) A victim client is selected, and the information needed is now acquired to create the backdoor model (9) and send it to the victim client (Section~\ref{sec:backdoor}).

\subsection{Training the Network}
\label{sec:train}

The FL network is composed of clients whose datasets are either IID or Non-IID split. Each client in the network trains the same DL algorithm with their dataset, and shares the locally trained model with the aggregator. The attacker keeps a historical record of each anonymously submitted model and a representative at each aggregation. A representative is a simpler representation of the model, acquired by querying a holdout data piece through the convolutional layers of the model, getting the embedded space before the softmax layer. Different inputs create different embedded spaces, so the input is fixed, achieving more stable representatives of the models for different clients and epochs~\cite{song2020analyzing}.

Finally, the server aggregates each update by FedAvg~\cite{mcmahan2017communication} and submits the joined model back to clients. This procedure is repeated until convergence is met. See Algorithm~\ref{alg:networktraining} for a summary.

% We compose the FL network with clients with specific \todo{specific how?} non-balanced labeled data. For aggregation, the client performs local training over its dataset and submits the anonymous model weights to the server. Anonymization can be achieved by different means, e.g., using TOR as suggested in~\cite{song2020analyzing}.\todo{it is not clear do we do anonymization or just discuss it?} 
% For every FL epoch $t$, the attacker extracts the representatives of each client uploaded model by querying a single holdout data piece and getting the inner computations of the second last layer (the layer before the fully connected layer), as in~\cite{song2020analyzing}.\todo{this needs to be better explained. I think I already told you this multiple times, and you still did not do it. You cannot just say we do it as someone else, and that is all. Now, the reader first needs to understand that other paper before being able to follow our attack}

% Since we want to maximize the representatives' resemblance for the same client's model for every $t$, we fix the queried sample, so the alteration of the model over $t$ would be slighter.\todo{not clear}
% Then, the server aggregates each update by FedAvg~\cite{mcmahan2017communication} and submits the joined model back to clients. This procedure is repeated until convergence is met. See Algorithm~\ref{alg:networktraining} for a summary.

\begin{algorithm}
\small
\caption{FL Network Training}
\label{alg:networktraining}
\begin{algorithmic}[1]
\State \textbf{Input:} A set of clients $K$. The number of clients $N$. Number of epochs $T$. Client local model $u$. Global model $W$.
\State \textbf{Output:} A collection of anonymous clients' representatives $D_{rep}$. A collection of anonymous clients' models per epoch $M$. 
\State \textbf{Initialize:} $W$
\State $\textbf{x} \leftarrow$ get\_sample() \Comment{Get the fixed sample.}
\For {each epoch $t=1,2,3,...,T$}
    \For {each client $k\in K$} \Comment{$k$ is anonymous for the server.}
        \State $u_{t+1}^k \leftarrow client\_update(k, W_t)$ \Comment{Local training of $k$.}
        \State $D_{rep\ t+1}^k \leftarrow u_{t+1}^k(\textbf{x})$ \Comment{Representatives over input $\textbf{x}$}
        \State $M_{t+1}^k \leftarrow u_{t+1}^k$
    \EndFor
    \State $W_{t+1} \leftarrow W_t + \frac{1}{N} \sum_{i=1}^{N} u_{t+1}^i$
\EndFor
\end{algorithmic}
\end{algorithm}

\subsection{Creating Synthetic Data}
\label{sec:creatingData}

Each party in the shadow network replicates the behavior of the regular FL procedure. Therefore, each shadow client requires a dataset similar to the one of the real clients. Model inversion attacks reconstruct the dataset a model has been trained on. To this end, we develop a GAN-based model inversion attack, adapting prior work~\cite{song2020analyzing}. Recent work~\cite{chen2020beyond} suggested replacing the discriminator with the client model. Consequently, the GAN will generate data similar to that specific client. 

We experimentally noticed that models do not have their properties merged at early epochs. For example, a model trained on a specific dataset can classify the seen data correctly. However, after a certain number of epochs, the aggregation merges properties from different clients, allowing the prediction of data owned by another client. Consequently, clients' models are similar when the global model reaches convergence, which is essential when choosing a model for replacing the GAN's discriminator. To create data that is specific to each client, the discriminator should not have the merged properties. Therefore, selecting an early epoch model as the discriminator is necessary. Selecting the proper epoch $t$ is use-case dependent. For simpler models and datasets, a few epochs are enough to learn the properties of the own model without fully merging the rest. However, more complex setups could require more training. We observe that carefully choosing $t$ is essential for the Non-IID case. In IID settings, models are similar from the beginning; thus, this effect is almost invisible. For details, please refer to Section~\ref{sec:experiments}. The procedure is summarized in the Algorithm~\ref{alg:data augmentation}.
%\onote{This is a long paragraph, can we divide into two?}

% To train the shadow network, we first need to create a dataset similar to the clients', see Algorithm~\ref{alg:data augmentation}. As in~\cite{chen2020beyond,song2020analyzing}, we develop a deep convolutional GAN (DCGAN) where the discriminator is created by each client's model at epoch $t$. We modified the client updated model by removing the last fully connected layer by another convolution and a sigmoid activation function to fit the DCGAN training procedure, as in~\cite{song2020analyzing}. \todo{again, we need a better explanation}
% After training, the generated samples are labeled by the last aggregated model, which has the greatest accuracy once convergence is met. As the data are distributed non-IID, the models require several epochs to combine their properties, leading to more different models in the earlier epochs than in the later epochs. Choosing the right $t$ will depend on the use case and may yield different results. Therefore, intuitively, it is advisable to choose a low $t$ to facilitate later identification.\todo{do we have something more than intuition how someone should select t?}
% Since the images generated by the GAN will best suit each client, this effect will then be transferred to the shadow training and, consequently, ease the identification process.\todo{not clear}

\begin{algorithm}[!ht]
\small
\caption{Creating Synthetic Data}
\label{alg:data augmentation}
\begin{algorithmic}[1]
\State \textbf{Input:} $K$ set of clients. $M$ collection of anonymous clients' models. Trained global model $W'$. $T$ number of epochs. $D$ is the discriminator, and $D$ is the discriminator.
\State \textbf{Output:} $D_{GAN}$ collection of GAN generated datasets.
\State \textbf{Initialize:} $t=1$ \Comment{Set a low value of $t$.}
\For{each client $k=1,2,3,...,K$}
    \State \textbf{Initialize:} $G$ and $D$.
    \State $D \leftarrow W_t^{'k}$
    \For{each epoch $i=1,2,3,...,T$}
        \State $\textbf{z} \leftarrow generate\_noise()$
        \State $train(G, M, \textbf{z})$
    \EndFor
    \State $z \leftarrow generate\_noise()$
    \State $\textbf{x} \leftarrow G(\textbf{z})$ \Comment{Create fake data.}
    \State $\{\textbf{x},y\} \leftarrow W'(\textbf{x})$ \Comment{Label data.}
    \State $X^k \leftarrow \{\textbf{x},y\}$
\EndFor
\end{algorithmic}
\end{algorithm}

\subsection{Shadow Training}
\label{sec:shadow}

Shadow models were first introduced by Shokri et al.~\cite{shokri2017membership} for a membership inference attack. %that determines if a data record was present in the training dataset. 
We modify the base idea to fit the FL requirements. 
Under our settings, we replicate the entire FL network in an isolated environment, namely a shadow network. It is composed of shadow clients, models, datasets, and a server. The attacker gains white-box access to the shadow training procedure by shadowing the FL network. We emphasize that neither the actual clients, their datasets, nor the server is used in the shadow network. The shadow clients have the GAN-generated samples as their training set and locally train the same DL algorithm in the ``real'' FL procedure.

Similarly, we also extract the shadow representatives of the shadow models (which are not anonymous now) and keep a record of them for each shadow client and epoch. By shadow training, the attacker gains knowledge of the relationship between datasets and models in a similar way to the actual procedure. The procedure is summarized in Algorithm~\ref{alg:shadow}.

% Shadow models were first introduced by Shokri et al.~\cite{shokri2017membership} in an inference attack, determining if a data record was present in the training dataset. Shadow training is based on creating a replica of the original black-boxed training procedure, which an attacker cannot access. The attacker has white-box access to every parameter or information by shadowing the training procedure. In our proposal, inspired by~\cite{song2020analyzing}, we shadow the entire FL network to gain white-box access to the training procedure, see Algorithm~\ref{alg:shadow}.

\begin{algorithm}
\small
\caption{Shadow Training}
\label{alg:shadow}
\begin{algorithmic}[1]
\State \textbf{Input:} A set of GAN generated dataset $D_{GAN}$. Set of shadow clients $\hat{K}$. Number of epochs $T$. The number of shadow clients $\hat{N}$. $\hat{u}$ shadow client local model. $\hat{W}$ shadow global model.
\State \textbf{Output:} $\hat{D}_{rep}$ shadow clients' representatives dataset.
\State $\textbf{x} \leftarrow get\_sample()$ \Comment{Get the fixes sample for calculating clients' representatives.}

\State \textbf{Initialize:} $\hat{W}$
\For{each epoch $t=1,2,3,...,T$}
    \For{each client $\hat{k}\in \hat{K}$}
        \State $\hat{u}_{t+1}^k \leftarrow client\_update(\hat{k},\hat{W}_t,D_{GAN}^{\hat{k}})$
        \State $\hat{D}_{rep\ t+1}^{\hat{k}} \leftarrow \hat{u}_{t+1}^{\hat{k}}(\textbf{x})$
    \EndFor
    \State $\hat{W}_{t+1} \leftarrow \hat{W}_t + \frac{1}{\hat{N}} \sum_{i=1}^{\hat{N}} \hat{u}_{t+1}^i$
\EndFor
\end{algorithmic}
\end{algorithm}

% The training procedure is equal to the one introduced in Section~\ref{sec:train}. However, now the attacker knows the model submitted by the clients, i.e., models are now identified. Therefore, the attacker can get a model representative for each client and epoch. In summary, by shadow training, the attacker gains white-box access to the shadow training procedure and constructs a dataset of shadow model representatives and its shadow owner for every epoch.
% As already stated, each client trains the same model over the generated dataset. As in Section~\ref{sec:train}, the attacker calculates the clients' representatives at each epoch over now identified clients' models.\todo{?} 
% In summary, the attacker has created a dataset of identified clients' representatives by shadow training.

\subsection{SNN Training \& Update Identification}
\label{sec:SNN}
\label{sec:identification}

% In our attack, we obtain model representatives by extracting the latent space's inner computations before each client model's last layer by querying the same sample data piece. Since the data provided for training is complex, we require the usage of triplet mining to create Semi-Hard triplets, improving the quality of the model~\cite{schroff2015facenet}.

SNN has already been used for inference attacks, e.g.,~\cite{song2020analyzing}.
During the training of the SNN, the authors used two representatives of different models as input. The SNN seeks similarities between them and outputs a value between ``0'' and ``1'', where zero means very similar. To improve SNN utility, we adopt triplet networks for model inversion attacks. 

The attacker owns a record of anonymous model updates from the real network and a copy of not anonymous updates from the shadow network. To identify the unlabeled dataset, the attacker aims to find relations between the labeled dataset. Since data is multidimensional, we use the triplet SNN. Triplet SNN requires three inputs for training, which we construct using online triplet mining. By doing so, the SNN can measure the similarity of two given representatives. The attacker then uses the trained SNN to match a client's identity with its representative. 

Before injecting the backdoor, it is first necessary to identify the victim. Clients' updates are dissimilar at the first epochs, and they get similar as the aggregation process is repeated---a model from the first epoch and another one from the last are highly different. We develop a client identification algorithm, greedy searching for the closest relation between representatives across epochs, linking models at early and last epochs. We leverage the trained SNN and the anonymous set of model representatives acquired in Section~\ref{sec:train}, which stores anonymous representations of every client model at every epoch. 
Precisely, a representative is chosen $rep_{t=0}^i$ at the first epoch $t=0$, and compared with another one in the subsequent epoch $t=1: \{0,1\} \leftarrow SNN(rep_{t=0}^i,rep_{t=1}^j)$. If the representatives are similar $1\simeq SNN(rep_{t=0}^i,rep_{t=1}^j)$, then it belongs to the same client $i=j$, otherwise another representative is chosen from the same epoch. The process is repeated for all the consecutive epochs in training.
After the process finishes, the algorithm maps the representatives across epochs, which is used to identify the clients. Furthermore, since the generated dataset is created from a client model, this process allows the mapping between the dataset and the client, easing selecting a victim.

\subsection{Backdoor Attack}
\label{sec:backdoor}

\begin{figure}[!ht]
     \centering
     \begin{subfigure}[b]{0.15\textwidth}
         \centering
         \includegraphics[width=\textwidth]{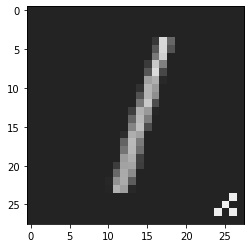}
         \caption{MNIST}
     \end{subfigure}
     %\hfill
     \begin{subfigure}[b]{0.15\textwidth}
         \centering
         \includegraphics[width=\textwidth]{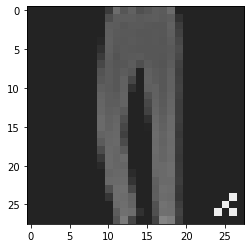}
         \caption{F-MNIST}
     \end{subfigure}
     %\hfill
     \begin{subfigure}[b]{0.15\textwidth}
         \centering
         \includegraphics[width=\textwidth]{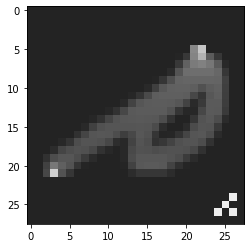}
         \caption{EMNIST}
     \end{subfigure}
     %\hfill
    \caption{Backdoored images.}
    \label{fig:backdoor_image}
\end{figure}

Backdoor attacks directly inject the adversarial effect on the model, fired by a trigger in the input sample. 
The backdoor is created by injecting a four-pixel pattern in the bottom right corner into some data pieces (see Figure~\ref{fig:backdoor_image}). Different trigger types, positioning, and combinations lead to variations in the clean data and backdoor accuracy, which we omit here. We refer the reader to~\cite{gu2019badnets} for the details of the effectiveness of different triggers. 

In ML and FL, backdoors poison the dataset for several epochs, starting from the first one~\cite{gu2019badnets}. However, our proposal injects the backdoor by retraining the model for a few epochs, which is then sent to the victim client, while the rest receive the non-poisoned one, see Algorithm~\ref{alg:backdoor}. The backdoor gets injected at near convergence to boost the backdoor performance, which uses a better trained model as a baseline, and to fit the necessities of our attack.
Since the training stops due to stopping criteria, e.g., the loss does not change more than 10\% for five epochs, the attacker sets a value $\lambda$ to anticipate the stopping criteria---injecting the backdoor in an almost optimal model. The backdoor is inserted in the model at an epoch that satisfies $\lambda\pm$ the stopping criteria. A small $\lambda$ ensures that the model is close to convergence, i.e., better model accuracy. Further consideration of the defense evasions is provided in Section~\ref{sec:defense}.

\begin{algorithm}
\small
\caption{Client identification \& Backdoor}
\label{alg:backdoor}
\begin{algorithmic}[1]
\State \textbf{Input:} Unidentified clients' representatives $D_{rep}$. Target class $c_t$. Source class $c_s$. GAN generated datasets $D_{GAN}$. Poisoned data rate $\epsilon$.
\State \textbf{Output:} Backdoored model $W_{\hat{\theta}}$.
\For{each unidentified client representative pair $x,x' \in D_{rep}:x\neq y$}
    \State $SNN(x,x')$ \Comment{Similarity calculation as in Section~\ref{sec:identification}}
\EndFor
\State \textbf{Define:} $u_{v}$ \Comment{Define a victim client}
\State $\hat{D}_{train} \leftarrow backdoor(c_s,c_t,D_{GAN}^v,\epsilon)$
\State $W_{\hat{\theta}} \leftarrow train(\hat{D}_{train},u_v)$
\State Send $W_{\hat{\theta}}$ to victim client $v$. 
\end{algorithmic}
\end{algorithm}

%\todo{it looks to me that a lot of threat model is actually given here?}

\section{Experimental Results}
\label{sec:experiments}

In this section, we provide an overview of the evaluated datasets and experimental setup, followed by the results and discussion.

\subsection{Datasets}

We evaluate the performance of our attack on the MNIST~\cite{lecun1998mnist}, EMNIST~\cite{cohen2017emnist}, and F-MNIST~\cite{xiao2017fashion} datasets. {MNIST} is a common benchmark dataset in computer vision containing labeled grayscale images from handwritten digits. Dataset labels range from ``0'' to ``9''. {EMNIST} is a grayscale dataset containing handwritten characters of the alphabet containing 26 classes of images. {F-MNIST} is a grayscale dataset containing ten types of clothing. Every dataset contains 70\,000 28$\times$28$\times$1 grayscale samples, 60\,000 for training, and 10\,000 for the test set.
Our selection of datasets allows us to consider standard settings and investigate scenarios with different number classes.

\subsection{FL Network Settings}
\label{sec:fl_settings}
For each dataset, the model is a Convolutional Neural Network (CNN), with three convolutional layers and a fully connected one, with stochastic gradient descent, LeakyRelu as an activation function, and batch normalization in each layer except the last. Experimentally set training settings are shown in Table~\ref{tab:settings}. The architecture is shown in Table~\ref{tab:cnn_arch}, and it is a commonly used convolution network for image classification tasks~\cite{szegedy2015going}.

\begin{table}[!htb]
\centering
\caption{Training settings.}
\begin{tabular}{P{1.2cm}P{0.8cm}P{1.2cm}P{0.6cm}P{0.6cm}P{0.6cm}P{0.6cm}}
\hline
Dataset &
  \textit{LR}&
  \begin{tabular}[c]{@{}c@{}} Momentum \end{tabular} &
  \begin{tabular}[c]{@{}c@{}}Local\\ Epoch\end{tabular} &
  \begin{tabular}[c]{@{}c@{}}FL \\ Epoch\end{tabular} &
  \begin{tabular}[c]{@{}c@{}}No. of\\ Clients\end{tabular} &
  \begin{tabular}[c]{@{}c@{}}No. of\\ Classes\end{tabular} \\ \hline 
MNIST   & 0.1     & 0.8 & 2 & 50  & 5  & 10 \\ 
F-MNIST & 0.00001 & 0.2 & 1 & 200 & 5  & 10 \\ 
EMNIST  & 0.01    & 0.8 & 2 & 200 & 13 & 26 \\ 
EMNIST~(IID) & 0.01    & 0.8 & 2 & 30 & 13 & 26 \\ 
\hline
\end{tabular}
\label{tab:settings}
\end{table}

\begin{table}[!htb]
\centering
\caption{CNN architecture.}
\label{tab:cnn_arch}
\begin{tabular}{@{}ccc@{}}
\toprule
Layer       & Out Shape          & Param \# \\ \midrule
Conv2D      & (None, 64, 14, 14) & 640      \\
LeakyReLu   & (None, 64, 14, 14) & -         \\
Conv2D      & (None, 128, 7, 7)  & 73\,856  \\
BatchNorm2D & (None, 128, 7, 7)  & 256      \\
LeakyReLu   & (None, 128, 7, 7)  & -        \\
Conv2D      & (None, 256, 3, 3)  & 295\,168 \\
BatchNorm2D & (None, 256, 3, 3)  & 512      \\
LeakyReLu   & (None, 256, 3, 3)  & -        \\
Linear      & (None, $10^\dag$)        & 23\,050  \\ \midrule
Total       &                    & 393\,482 \\ \bottomrule
\end{tabular}%
\\
\footnotesize{$^\dag$ ``10'' changes with the number of classes.}
\end{table}

\begin{figure*}[!ht]
\centering
    \begin{subfigure}[b]{0.33\textwidth}
        \centering
        \includegraphics[width=\linewidth]{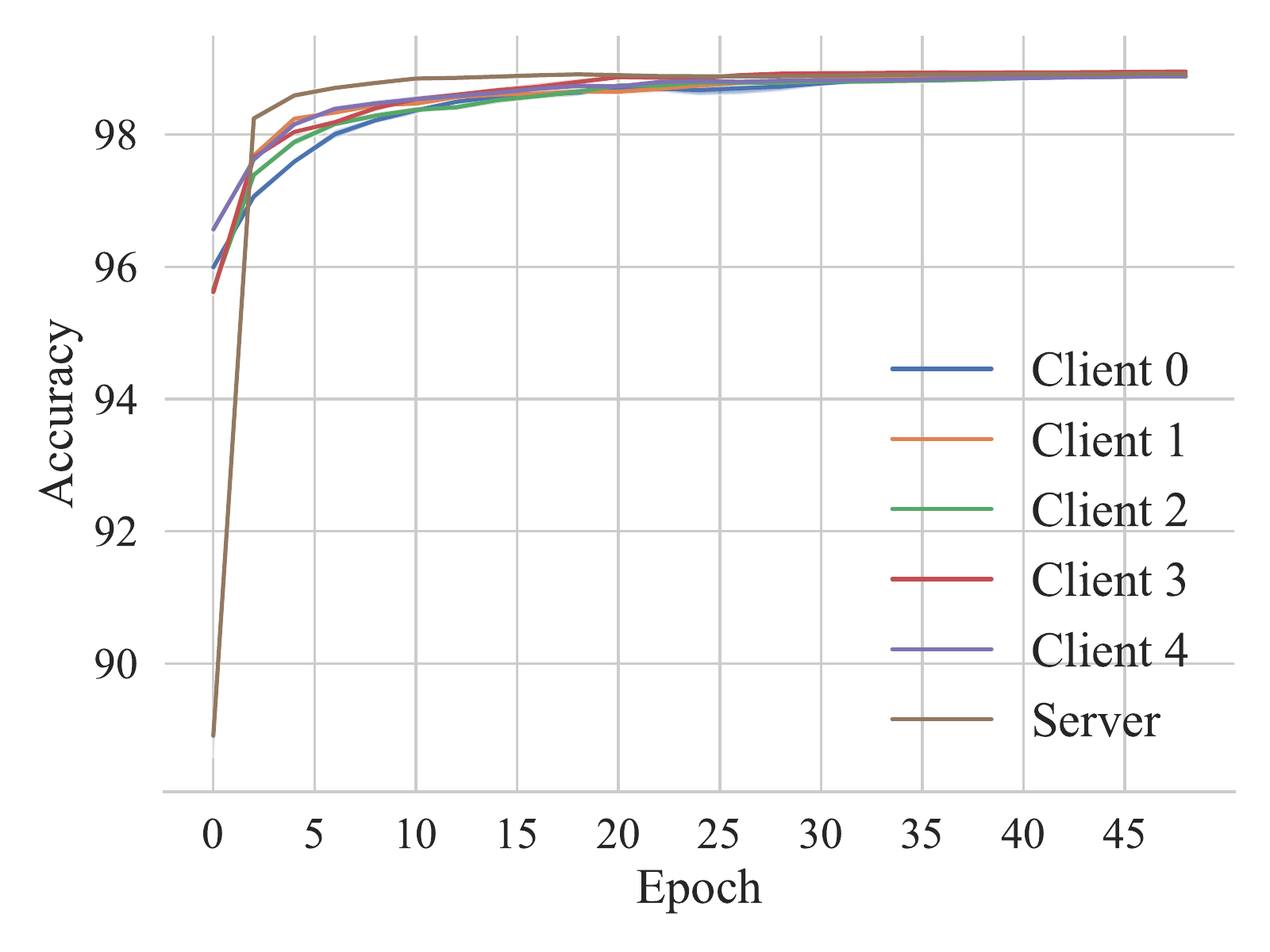}
        \caption{MNIST test accuracy with IID data.}
        \label{fig:mnist_iid}
    \end{subfigure}
\hfill
    \begin{subfigure}[b]{0.33\textwidth}
        \centering
        \includegraphics[width=\linewidth]{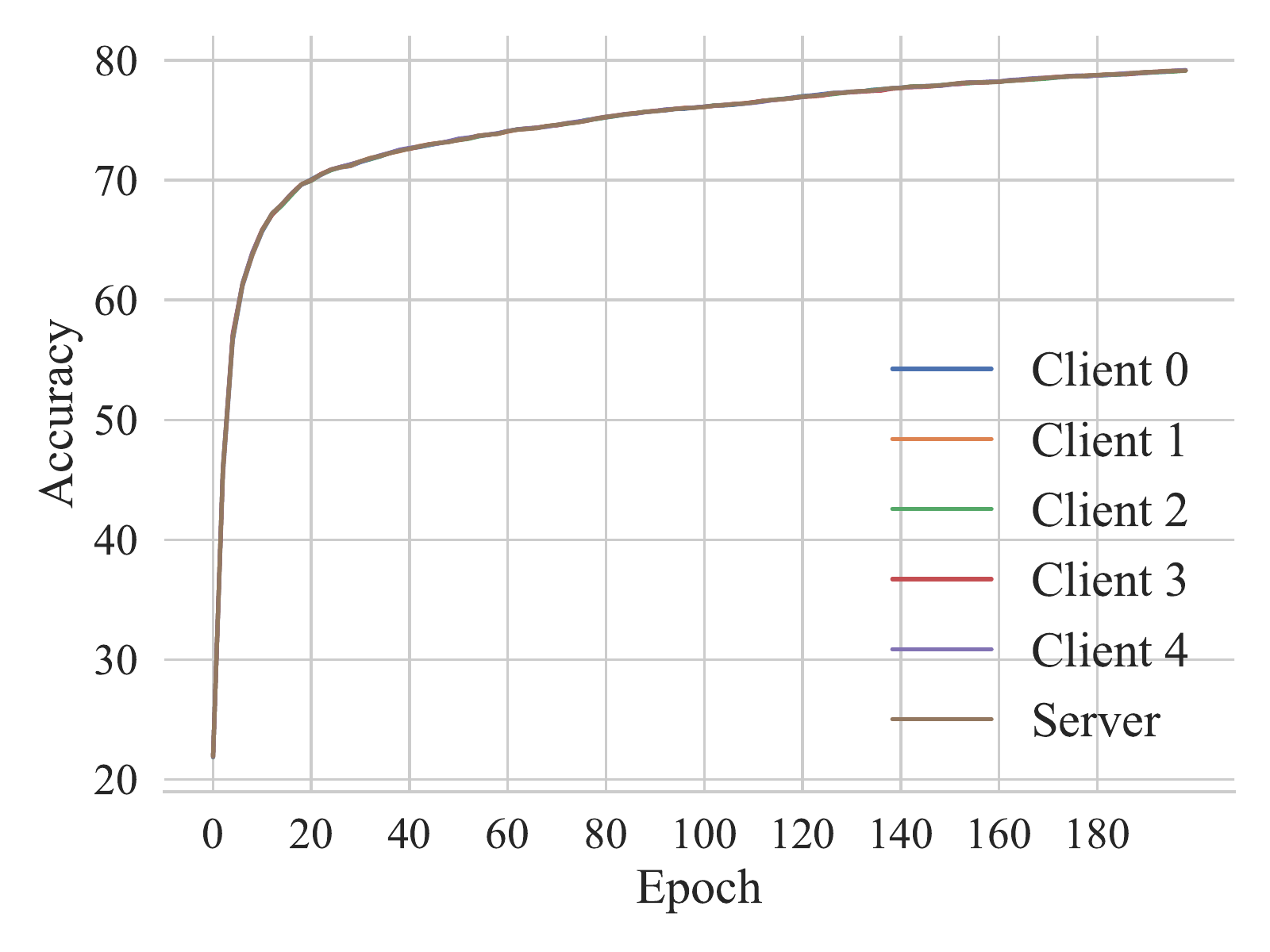}
        \caption{F-MNIST test accuracy with IID data.}
        \label{fig:fmnist_iid}
    \end{subfigure}
\hfill
    \begin{subfigure}[b]{0.32\textwidth}
        \centering
        \includegraphics[width=\linewidth]{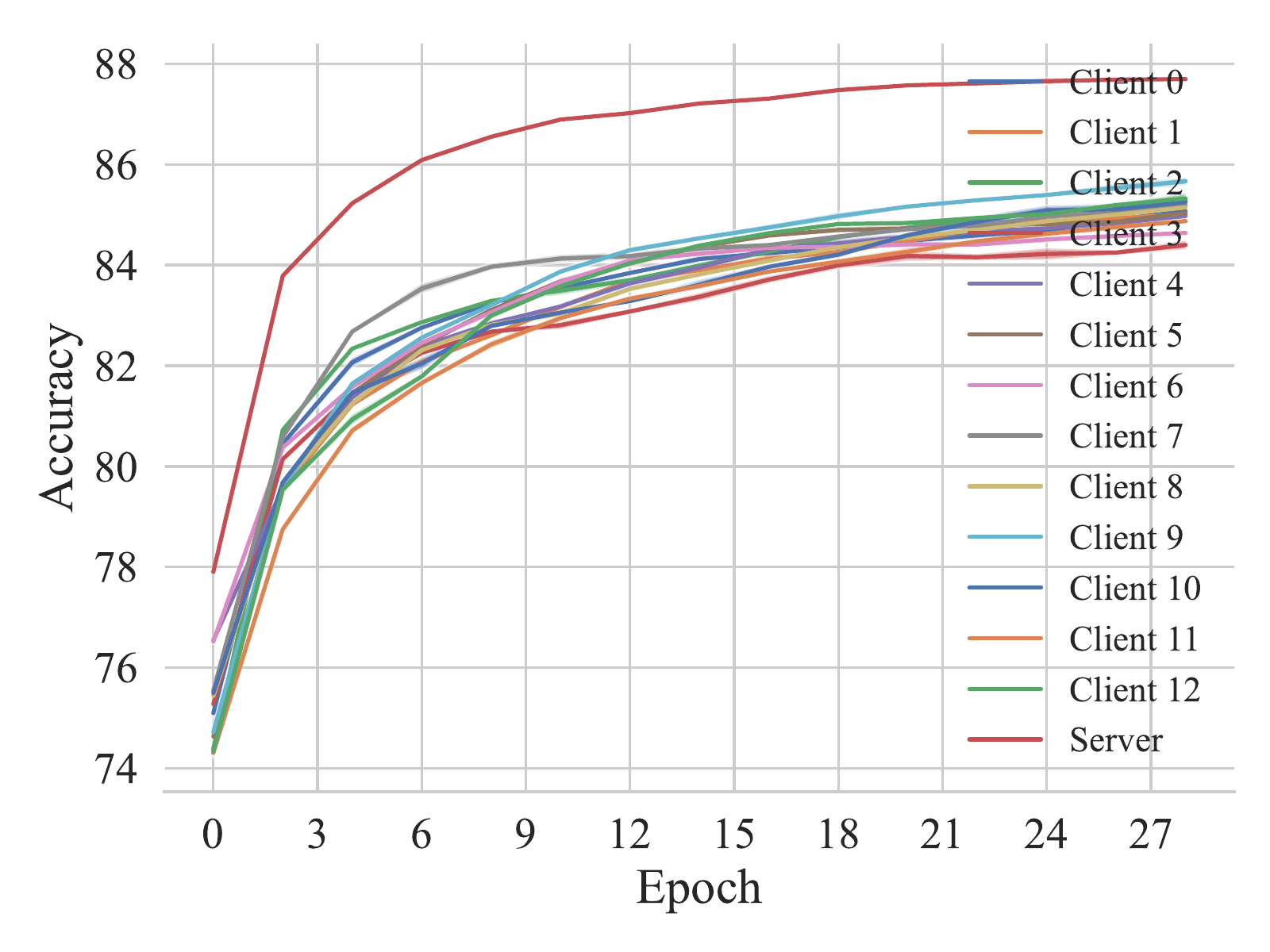}
        \caption{EMNIST test accuracy with IID data.}
        \label{fig:emnist_iid}
    \end{subfigure}
    \begin{subfigure}[b]{0.33\textwidth}
        \centering
        \includegraphics[width=\linewidth]{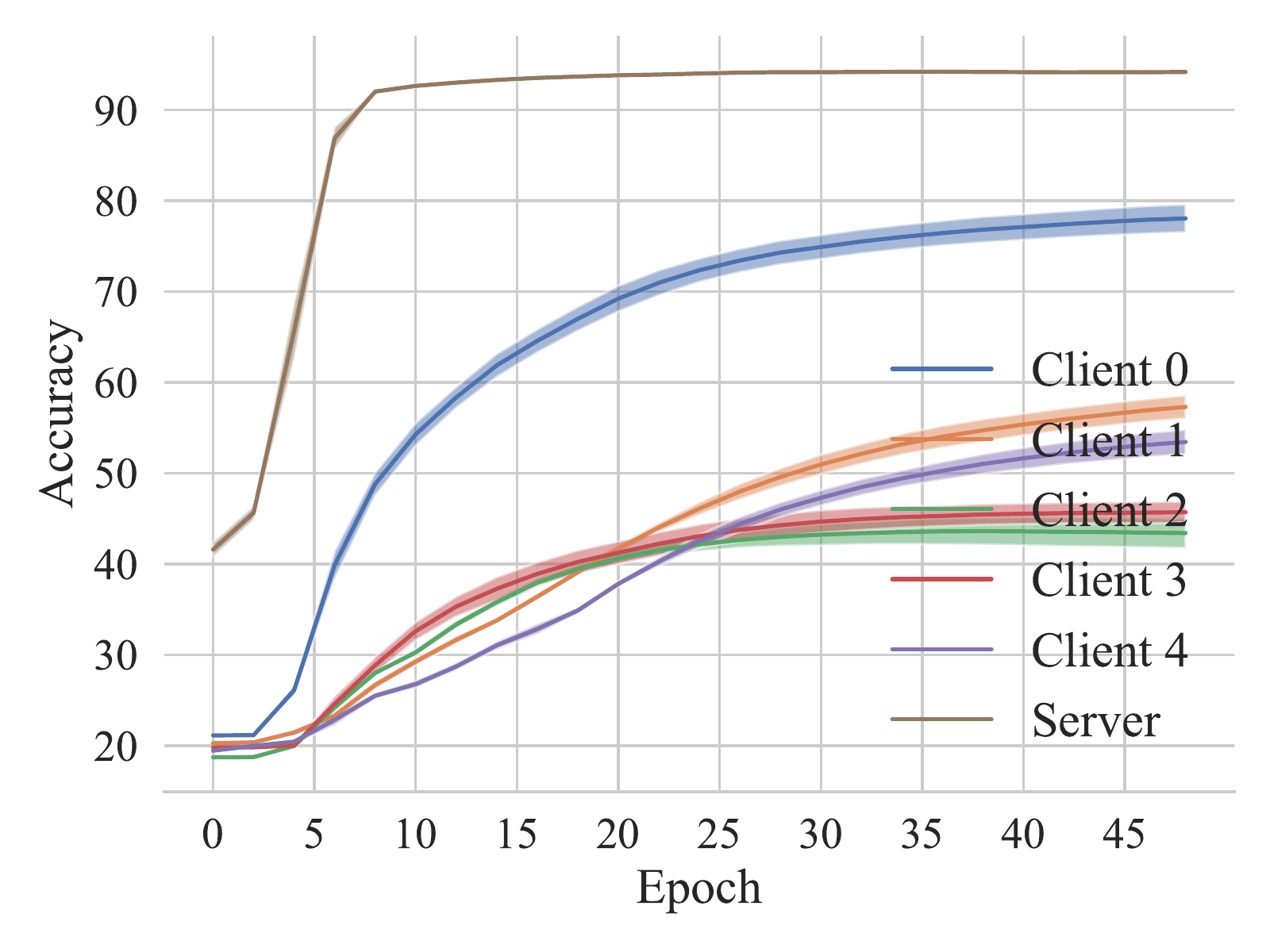}
        \caption{MNIST test accuracy with Non-IID data.}
        \label{fig:mnist_non_iid}
    \end{subfigure}
\hfill
    \begin{subfigure}[b]{0.33\textwidth}
        \centering
        \includegraphics[width=\linewidth]{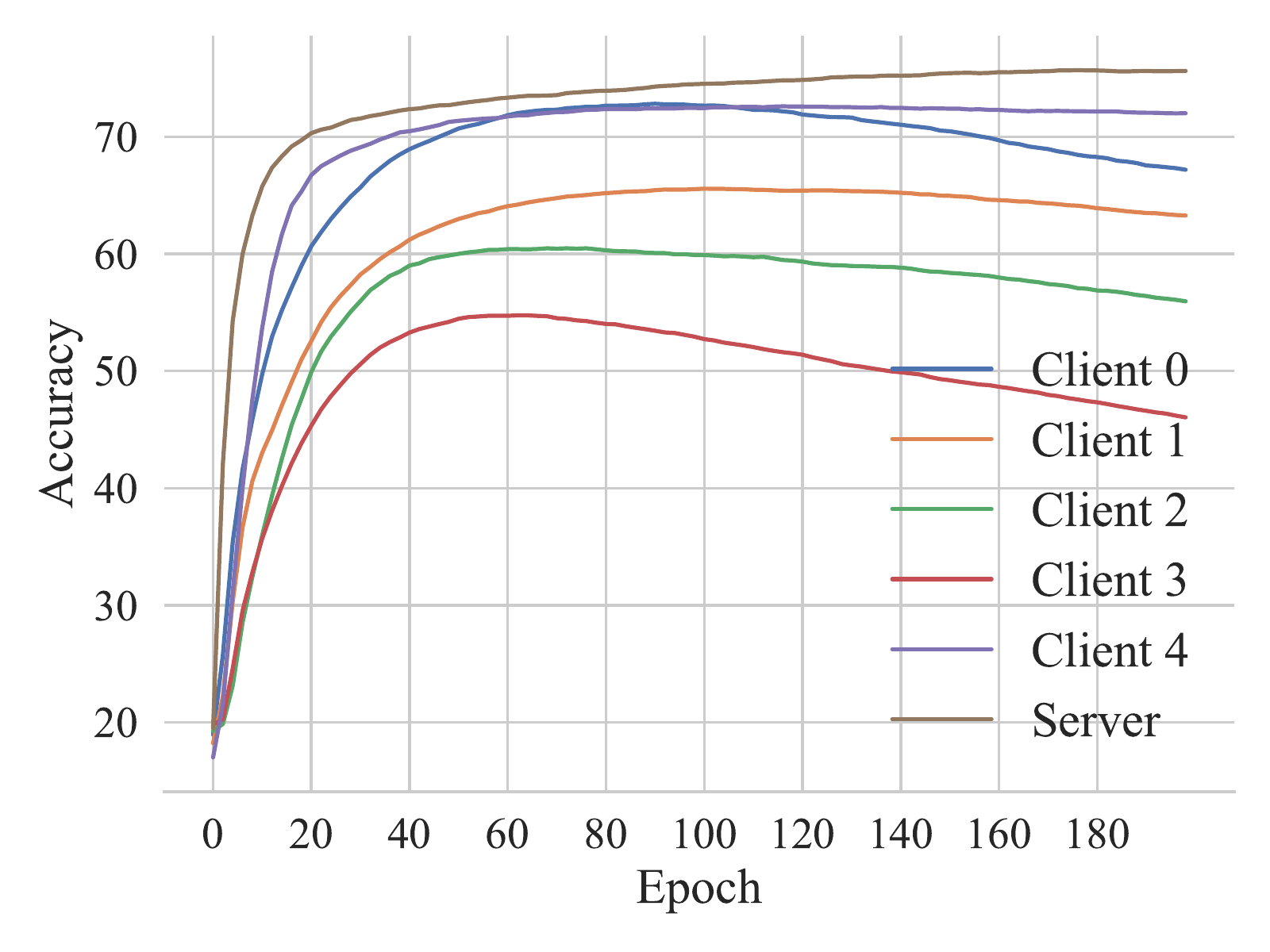}
        \caption{F-MNIST test accuracy with Non-IID data.}
        \label{fig:fmnist_non_iid}
    \end{subfigure}
\hfill
    \begin{subfigure}[b]{0.32\textwidth}
        \centering
        \includegraphics[width=\linewidth]{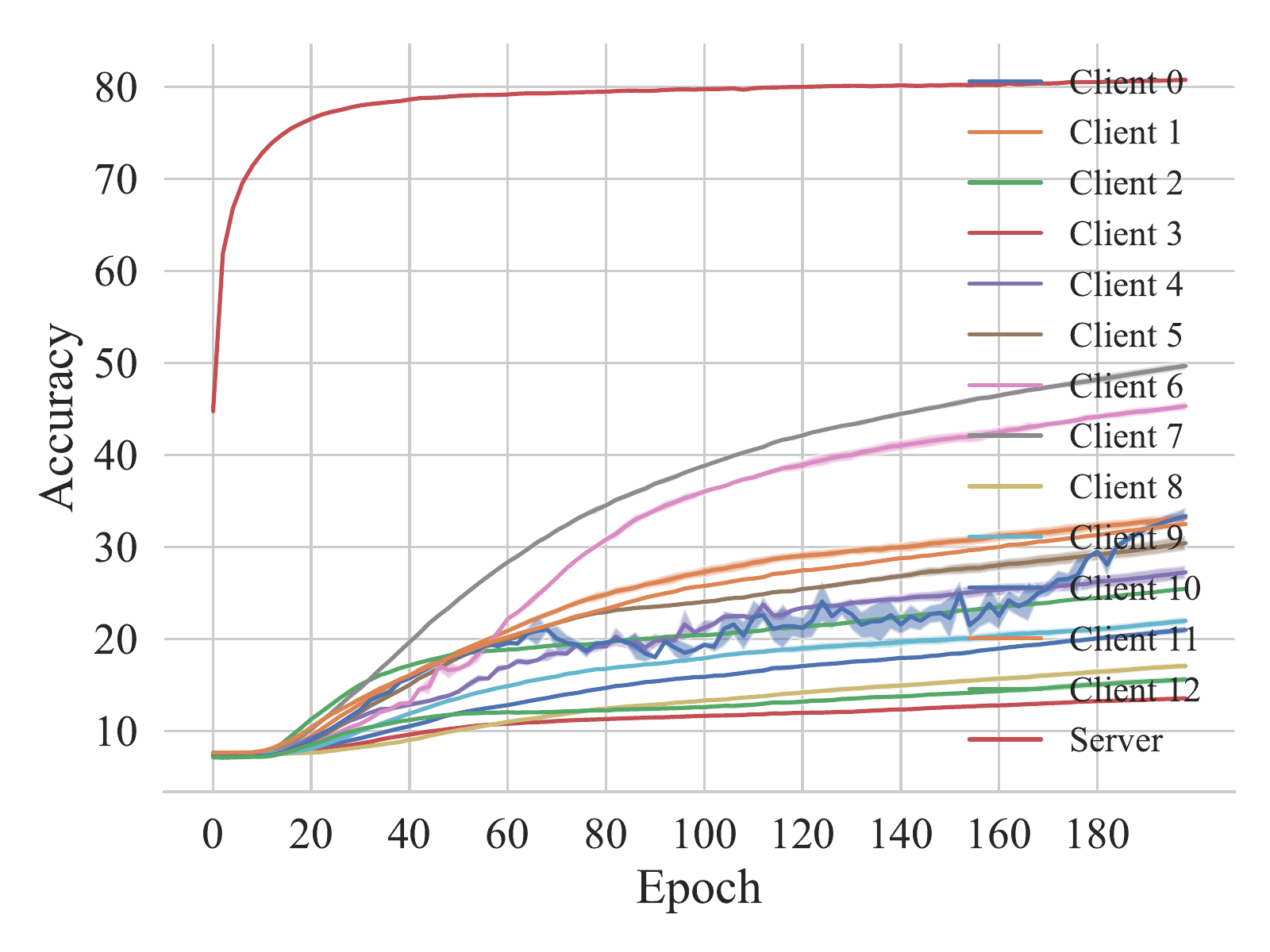}
        \caption{EMNIST test accuracy with Non-IID data.}
        \label{fig:emnist_non_iid}
    \end{subfigure}
\caption{CNN averaged testing accuracy for different datasets.}
\label{fig:traning_avg}
\end{figure*}

After training, with the Non-IID setting, the network achieves 95\% accuracy on MNIST, 78\% on F-MNIST, and 80\% on EMNIST, see Figure~\ref{fig:traning_avg}.\footnote{Results are averaged over ten executions.} Regarding the IID setting, the model reaches better results, 99\% on MNIST, 80\% on F-MNIST, and 88\% on EMNIST. Figure~\ref{fig:traning_avg} shows the test accuracy of the global model (server) and the clients' model after local training. Models are trained over non-colluding (Non-IID) or overlapping (IID) labeled data and evaluated with a test dataset containing all the labels. As epochs progress, models perform better over the test set, acquiring properties from other datasets.

\begin{figure*}[ht]
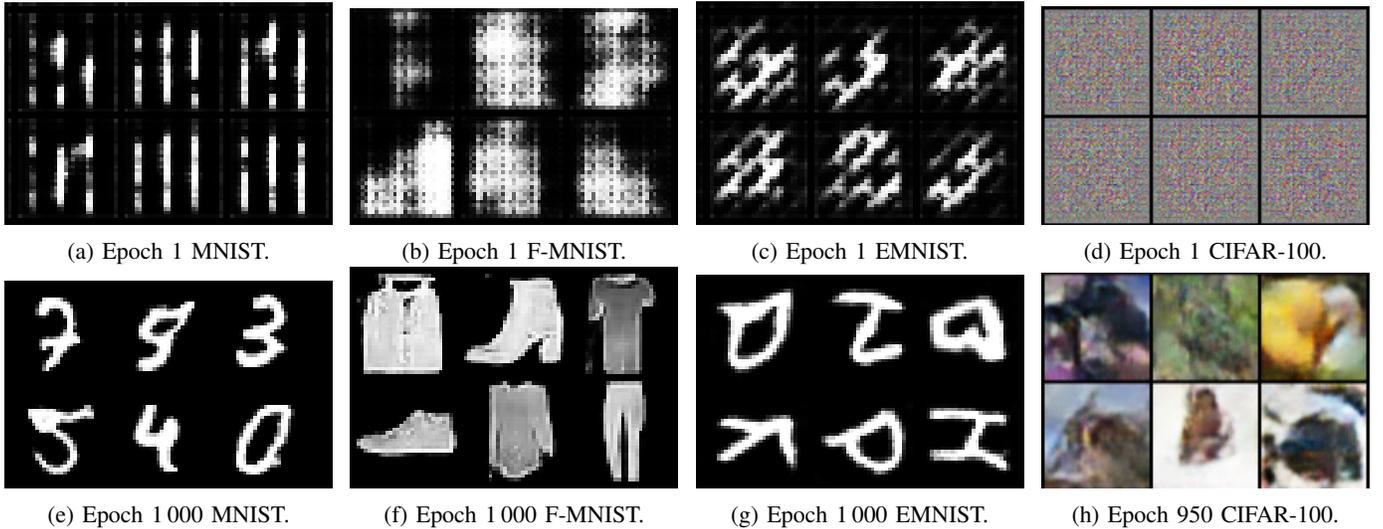

     \centering
     \begin{subfigure}[b]{0.32\textwidth}
         \centering
         \includegraphics[width=\textwidth]{figures/fake_images-1_reduced.png}
         \caption{Epoch 1 MNIST.}
     \end{subfigure}
     \hfill
     \begin{subfigure}[b]{0.32\textwidth}
         \centering
         \includegraphics[width=\textwidth]{figures/FMNIST_fake_1.png}
         \caption{Epoch 1 F-MNIST.}
     \end{subfigure}
     \hfill
     \begin{subfigure}[b]{0.32\textwidth}
         \centering
         \includegraphics[width=\textwidth]{figures/EMNIST_fake_1.png}
         \caption{Epoch 1 EMNIST.}
     \end{subfigure}
     \hfill
     \begin{subfigure}[b]{0.32\textwidth}
         \centering
         \includegraphics[width=\textwidth]{figures/fake_images-2-200_reduced.png}
         \caption{Epoch 1\,000 MNIST.}
    \end{subfigure}
         \hfill
     \begin{subfigure}[b]{0.32\textwidth}
         \centering
         \includegraphics[width=\textwidth]{figures/FMNIST_fake_200.png}
         \caption{Epoch 1\,000 F-MNIST.}
     \end{subfigure}
     \hfill
     \begin{subfigure}[b]{0.32\textwidth}
         \centering
         \includegraphics[width=\textwidth]{figures/EMNIST_fake_1000.png}
         \caption{Epoch 1\,000 EMNIST.}
    \end{subfigure}
        \caption{GAN generated MNIST, F-MNIST, and EMNIST images at different epochs.}
        \label{fig:gan_images}
\end{figure*}

Regarding GAN, it is constructed by adapting the clients' models as the discriminator. The last fully-connected layer of the model has to be changed with a convolutional layer and sigmoid activation function, as in~\cite{song2020analyzing}, to fit the discriminator architecture. The generator is, otherwise, a concatenation of deconvolutional networks that upscale a vector of noise to a $28\times28$ figure. The entire architecture is presented in Tables~\ref{tab:discriminator} and~\ref{tab:generator}. We train GAN for 1\,000 epochs and experimentally set an \textit{LR} of 0.002 and Adam as the optimizer. The attacker generates 5\,000 labeled images per client (Figure~\ref{fig:gan_images}) that would be used in the shadow training.

\begin{table}[htb]
\centering
\caption{Discriminator architecture.}
\label{tab:discriminator}
\begin{tabular}{@{}ccc@{}}
\toprule
Layer       & Out Shape          & Param \# \\ \midrule
Conv2D      & (None, 64, 14, 14) & 640      \\
LeakyReLu   & (None, 64, 14, 14) &          \\
Conv2D      & (None, 128, 7, 7)  & 73\,856  \\
BatchNorm2D & (None, 128, 7, 7)  & 256      \\
LeakyReLu   & (None, 128, 7, 7)  & -        \\
Conv2D      & (None, 256, 3, 3)  & 295\,168 \\
BatchNorm2D & (None, 256, 3, 3)  & 512      \\
LeakyReLu   & (None, 256, 3, 3)  & -        \\
BatchNorm2D & (None, 1, 1, 1)    & 23\,004  \\
Sigmoid     & (None, 1, 1, 1)    & -        \\ \midrule
Total       &                    & 372\,736 \\ \bottomrule
\end{tabular}%
\end{table}

% \usepackage{booktabs}
% \usepackage{graphicx}
\begin{table}[htb]
\centering
\caption{Generator architecture.}
\label{tab:generator}
\begin{tabular}{@{}ccc@{}}
\toprule
Layer           & Out Shape           & Param \# \\ \midrule
ConvTranspose2D & (None, 256, 64, 64) & 409\,600 \\
BatchNorm2D     & (None, 256, 64, 64) & 512      \\
ReLu            & (None, 256, 64, 64) & -        \\
ConvTranspose2D & (None, 128, 7, 7)   & 294\,912 \\
BatchNorm2D     & (None, 128, 7, 7)   & 256      \\
ReLu            & (None, 128, 7, 7)   & -        \\
ConvTranspose2D & (None, 64, 14, 14)  & 131\,072 \\
BatchNorm2D     & (None, 64, 14, 14)  & 128      \\
ReLu            & (None, 64, 14, 14)  & -        \\
ConvTranspose2D & (None, 1, 28, 28)   & 1\,024   \\
Tanh            & (None, 1, 28, 28)   & -        \\ \midrule
Total           &                     & 837\,504 \\ \bottomrule
\end{tabular}%
\end{table}

% \begin{figure}[ht]
%      \centering
%      \begin{subfigure}[b]{0.31\textwidth}
%          \centering
%          \includegraphics[width=\textwidth]{figures/FMNIST_fake_1.png}
%          \caption{Epoch 1.}
%      \end{subfigure}
%      \hfill
%      \begin{subfigure}[b]{0.31\textwidth}
%          \centering
%          \includegraphics[width=\textwidth]{figures/FMNIST_fake_100.png}
%          \caption{Epoch 100.}
%      \end{subfigure}
%      \hfill
%      \begin{subfigure}[b]{0.31\textwidth}
%          \centering
%          \includegraphics[width=\textwidth]{figures/FMNIST_fake_200.png}
%          \caption{Epoch 200.}
%      \end{subfigure}
%         \caption{GAN generated F-MNIST images at different epochs.}
%         \label{fig:gan_images_fmnist}
% \end{figure}

\subsection{Shadow Network Training Settings}

A shadow network replicates the original FL network introduced in Section~\ref{sec:train} and utilizes the same hyperparameters as in Section~\ref{sec:fl_settings}. The shadow network contains shadow clients and shadow datasets. Shadowing the FL network allows the attacker to gain white-box access to the entire training procedure. Shadow clients own shadow datasets, i.e., the synthetic dataset generated previously containing 5\,000 data samples. Shadow clients train their shadow models and upload them to the shadow server. The attacker extracts each shadow model representative and matches them to a shadow client at every epoch. Since the shadow models are trained with synthetic data, their shadow models' accuracy is lower than the actual models. However, the shadow models' accuracy is not relevant, which are just used to extract identified representatives. After this process, the attacker owns a dataset of shadow model representatives.

% The shadow network is a replica of the original FL network, using the training parameters presented in Section~\ref{sec:train} but with the synthetic datasets. Each client owns a synthetic dataset created using the DCGAN from their models at an early epoch, e.g., $t \approx 1$. \todo{how does the client get the synthetic dataset, and why would he use it? Gorka: I think I already explained this in section 4.4}
% In the MNIST case, since the labeling is $\approx95\%$ accurate (the global model accuracy at convergence), errors in the dataset are introduced, lowering the shadow global model's accuracy to $\approx90\%$. In the F-MNIST case, the accuracy is $\approx76\%$ and $\approx89\%$ for EMNIST on average. However, shadow network accuracy is not relevant; its only assignment is to extract identified representatives. During the FL process, the attacker extracts the latent space of the second last layer over a fixed image as input, as in Section~\ref{sec:train}, to create a dataset for the SNN.\todo{DONE. what about other datasets?}

\subsection{Triplet SNN Training Settings}

Each model of the SNN is composed of three fully connected layers with dropout layers between them. This simple, yet effective design, grants excellent performance with low amounts of data. More complex models could easily overfit. Furthermore, we require triplet mining to improve the quality of the network. 
The network inputs are an anchor, a positive, and a negative 2\,304-dimensional samples. The online triplet mining procedure selects the best triplet combination, being the anchor and the positive sample representatives from the same client, while the negative is from another client.

The outputs from the last layers are embedded in five-dimensional space (experimentally set as a trade-off between network complexity and data dimensionality) and sent to a distance computing layer that calculates the Euclidean distance. After training, given two inputs, the SNN yields values close to 0 if they are similar and close to 1 otherwise. Our experiments show the network reaches an accuracy of 97\% and IID and 80\% with Non-IID for MNIST, 80\% and IID and 78\% with Non-IID for F-MNIST, 90\% and IID and 88\% with Non-IID for EMNIST, after 20 epochs, $\alpha$ 0.2, and an \textit{LR} of 0.0001, with Adam as the optimizer. We can observe a slight degradation in the SNN accuracy with IID data caused by the high models' resemblance.

\subsection{Backdoor Attack Settings}

The attacker poisons a dataset by injecting pixel pattern samples and flipping the source to the target label. The attacker uses the last joined model to inject the backdoor via training with poisoned data. A value $\epsilon$ controls the amount of poisoned data in the dataset.
A successful backdoor should preserve high accuracy on the source class while  maintaining high accuracy on the target class, measured by the metrics defined in Section~\ref{sec:advobjectives}. Note that a pixel pattern constructs the trigger as in BadNets~\cite{gu2019badnets}.
%Since this paper aims to establish the reality of client-targeted backdoors, the different types of triggers influence, e.g., single-pixel, are out of this paper's scope. We will research different triggers type and their influence in future work.

The attacker poisons a dataset by injecting a trigger on an input sample of a given source class and flips the label to the desired target class. The amount of samples containing the trigger is controlled by $\epsilon$. In our setting, only the source class samples are valid for the attack, which, combined with $\epsilon$, drastically lowers the valid poisoned sample candidates. Refer to Table~\ref{tab:epsilon_values} for the exact number of poisoned samples in the poisoned set. The trigger is a simple white 4-pixel pattern placed in the bottom right corner, as seen in Figure~\ref{fig:backdoor_image}. 

\begin{figure*}[!ht]
\centering
\includegraphics[width=\linewidth]{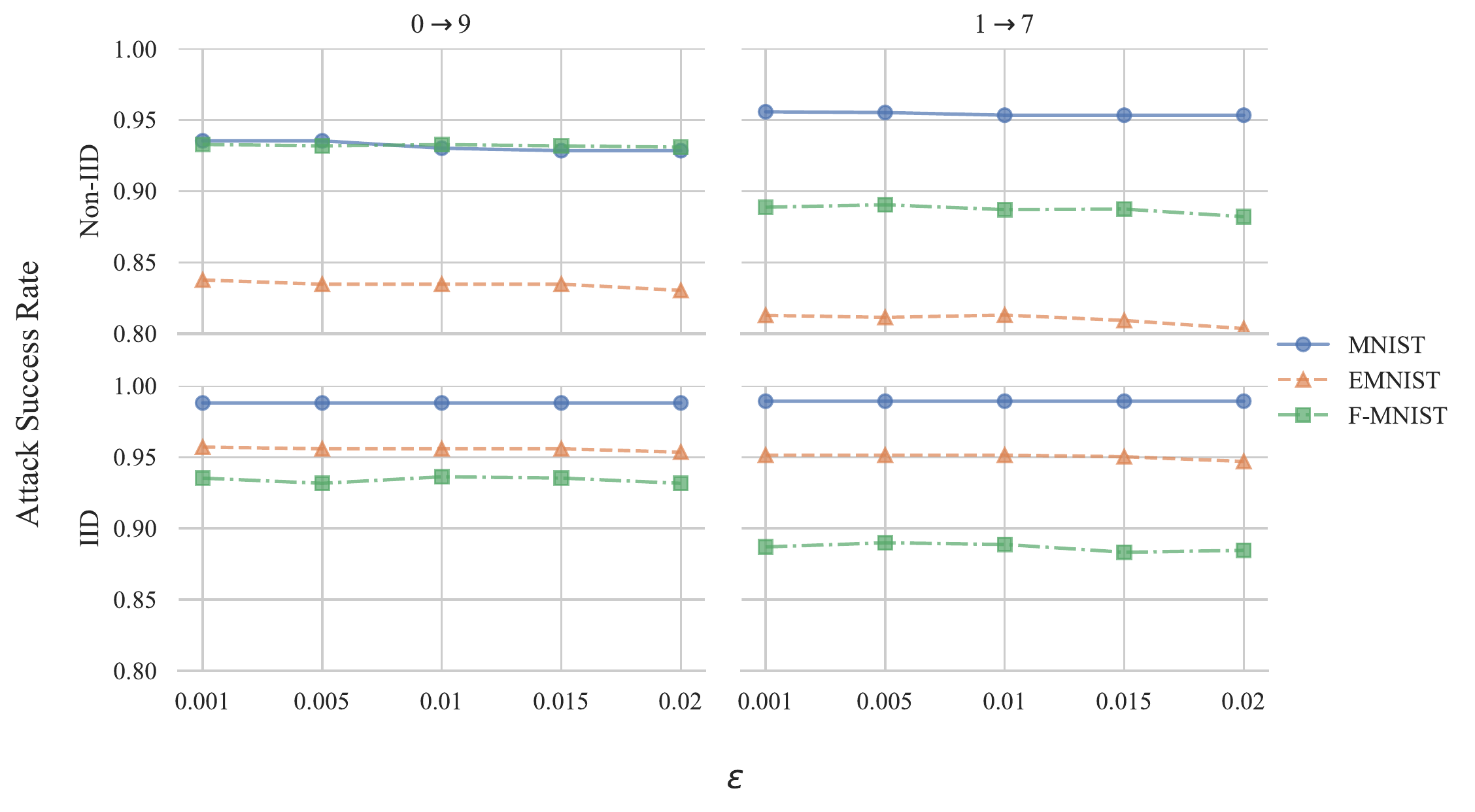}
\caption{ASR on the target class under different settings.}
\label{fig:backdoorASR}
\end{figure*}

\begin{figure*}[!ht]
\centering
\includegraphics[width=\linewidth]{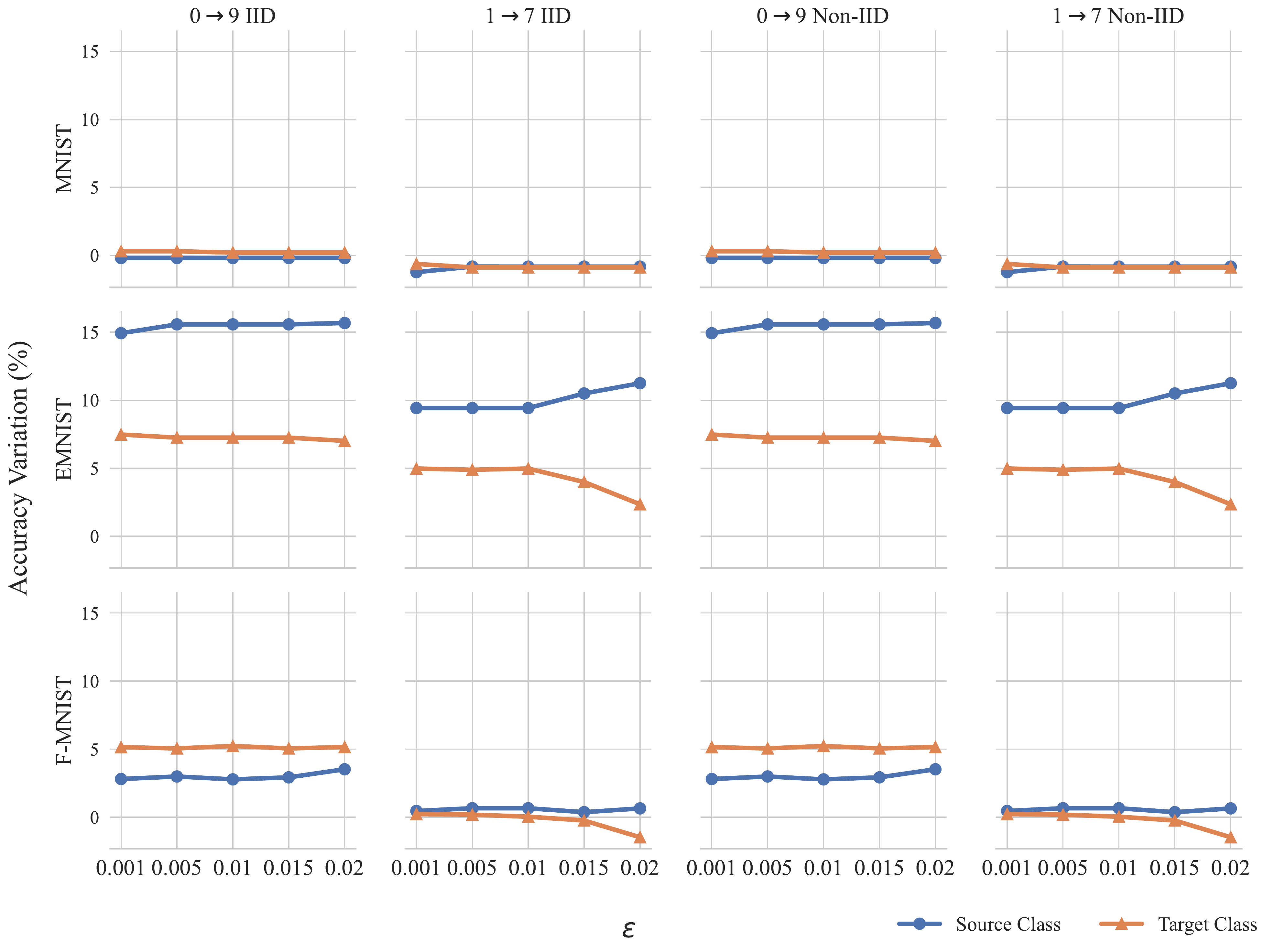}
\caption{Accuracy degradation (\%) on the target and source classes on a clean test set under different settings.}
\label{fig:backdoor_degradation}
\end{figure*}

We experiment with the attack by retraining the global model on the poisoned dataset for 10 epochs, LR 0.0001, SGD as the optimizer, and momentum 0.8 for MNIST and EMNIST for both IID and Non-IID settings. For F-MNIST, we retrain the model for 20 epochs, LR 0.01, SGD as an optimizer, and momentum of 0.8 for both IID and Non-IID. Using these hyperparameters, we define two attack scenarios: \textit{0 to 9}, where ``0'' is the source class and ``9'' is the target class, and \textit{1 to 7} where ``1'' is the source and ``7'' the target class, respectively for MNIST. In the F-MNIST dataset, ``0'' corresponds to ``T-shirt'', ``1'' to ``Trousers'', ``7'' to ``Sneakers'', and ``9'' to Ankle boot''. Regarding EMNIST, ``0'' corresponds to ``a'', ``1'' to ``b'', ``7'' to ``h'', and ``9'' to ``j''. We further evaluate the attack for different $\epsilon$ values and validate the attack performance by checking the target class ASR. We observe high ASR in every setting, see Figure~\ref{fig:backdoorASR}, and almost no degradation on either the source or target class concerning the accuracy of the main task (Figure~\ref{fig:backdoor_degradation}). However, the attacker should be careful when setting a large value of $\epsilon$, which could cause degradation on the main task while achieving a higher accuracy on the backdoor task. Furthermore, using a very large $\epsilon$ value $\epsilon=1$, i.e., no source class in the training set could cause the model to ``forget'' the source class, causing the model to perform poorly on that specific class. This finding could potentially be used to defend against such targeted attacks, as discussed in Section~\ref{sec:defense}. With the IID and Non-IID settings, the achieved ASR on the target class is up to 99\% in MNIST, 94\% in F-MNIST, and 96\% in EMNIST.

\begin{table}[htb]
\centering
\caption{The number of poisoned data samples in the dataset.}
\label{tab:epsilon_values}
\begin{tabular}{cccccc}
\hline
$\epsilon$    & 0.001 & 0.005 & 0.01 & 0.015 & 0.02 \\ \hline
\# of samples & 2     & 10    & 20   & 30    & 40   \\ \hline
\end{tabular}
\end{table}

\subsection{Comparison with the Existing Backdoor Attacks in FL}

This section compares our results with state-of-the-art backdoor attacks in FL. Before discussing the results, it is important to note that none of the existing methods target a single client, we try our best to find comparable results. 

Sun et al.~\cite{sun2019can} introduced backdoors attacks in FL, where they choose a subset of clients as attackers, which train their model on poisoned data. They use a CNN model and EMNIST as the dataset, which is similar to ours. Despite training with a different amount of clients, more epochs, and different LR, we still achieve similar results on the main and backdoor task, up to 80\% ASR, and less than 1\% degradation on the main task. Since backdoor models have a substantially larger norm than non-infected models, they test their attack against a norm threshold as a defense, drastically reducing the ASR while maintaining high accuracy on the main task.

Wang et al.~\cite{wang2020attack} investigated the same threat in FL in the image and text domains. The authors leverage \textit{edge case} samples---data pieces that rarely appear in the training dataset---as candidates for injecting the backdoor. They evaluate their attack against different existing techniques, which can bypass leveraging a projected gradient descent attack, clipping the norm of the model. Without defenses, despite the use of a different dataset (CIFAR-10 with some added edge case samples), they achieve comparable accuracy on the main (99\%) and backdoor task (80\%).

Lastly, \textit{DBA}~\cite{xie2019dba} is a distributed attack that separates the triggers into pieces and shares each with the attackers in the FL network. At the same epochs, the attackers will share the poisoned model---trained on a poisoned dataset containing the trigger piece---merging the trigger pieces at aggregation. The result is a fully backdoored model while stealthier than other centralized backdoor attacks. Their attack achieves 91\% ASR on the MNIST dataset, with the trigger placed at the upper left corner, similar to our approach.

\section{Defenses}
\label{sec:defense}

In this section, we explain the state-of-the-art defense mechanisms against backdoor attacks, and we evaluate and discuss their effectiveness and applicability against our attack.

\subsection{Generic Defense Methods}

To prevent backdoor attacks, several defense mechanisms have been developed recently. Specific to centralized ML, \textit{dataset inspection} techniques analyze the training set to remove outliners, assuming that the poisoned samples compose a small separate cluster~\cite{chen2018detecting}. Similarly, \textit{Neo}~\cite{udeshi2022model} generates different variants of the input sample, masking the dominant color, and checks the attack success rate for every pixel under every variant. The model is flagged as malicious if the attack success rate exceeds some threshold. 
\textit{STRIP}~\cite{gao2019strip} perturbs the samples by adding different patterns and observing the randomness of the model's outcome, which measures its entropy, and can detect if a model is poisoned or not. Dataset inspection defenses do not hold for FL settings, where the datasets are private; thus, we do not consider them for our experimentation.

\textit{Neural Cleanse}~\cite{wang2019neural} and \textit{ABS}~\cite{liu2019abs} defends against backdoor attacks by an optimization method for finding the smallest perturbation that makes the model behave abnormally. 
Since both defense mechanisms work similarly, in this work, we implement the Neural Cleanse defense mechanism and evaluate its effectiveness against our attack.
Neural Cleanse is based on the intuition that a poisoned model requires much less modification to cause misclassification into the trigger label than the rest. Iteratively, Neural Cleanse creates the same number of potential triggers as classes in the model, requiring minimal pixel changes to cause misclassification. Afterward, an outline detection algorithm selects the triggers smaller than the rest as malicious, assigning an anomaly score, and successfully reconstructing the target label and the trigger. Authors suggest using ``2'' as the threshold for finding malicious labels.

\begin{figure*}[htb]
    \centering
    \includegraphics[width=\linewidth]{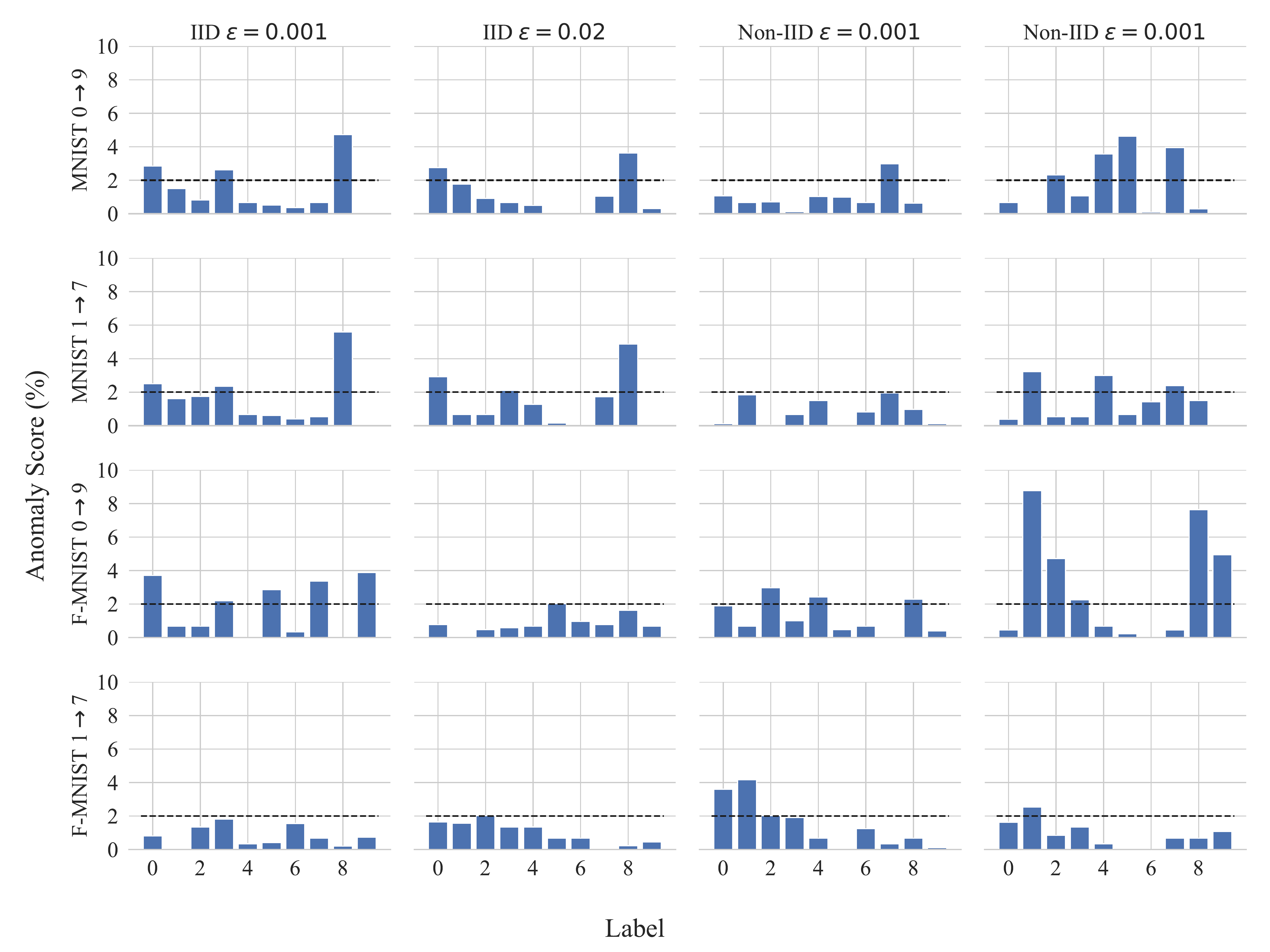}
     \caption{Anomaly detection of different attacks using Neural Cleanse~\cite{wang2019neural}.}
    \label{fig:neural_clease}
\end{figure*}

We implement this defense against our attack using \textit{Trojanzoo}~\cite{pang2022trojanzoo}, which has to be slightly modified for supporting source class targeted backdoors. We set up Neural Cleanse with the suggested hyperparameters and set it as default in the Trojanzoo implementation. We observe that Neural Cleanse cannot identify the target class as malicious, i.e., classes with an anomaly score greater than 2, in none of our attack settings (Figure~\ref{fig:neural_clease}). The attack reports primarily false positives, assigning high anomaly scores to several classes, sometimes even not including the targeted one. We observe that this effect is due to fixing the source class in our attack, given the intuition provided by the Neural Cleanse authors: \textit{``Our key intuition of detecting backdoors is that in an infected model, it requires much smaller modifications to cause misclassification into the target label than into other uninfected labels''}~\cite{wang2019neural}. In the attacks where the source class is not fixed, all the classes are converted to the target class, thus making the above statement true. However, when only a single class is used for firing the trigger, only small modifications of the source class inputs make it easier to cause misclassification into the targeted model. We can therefore conjecture that:
\textit{``In an infected model, detecting backdoors requires much smaller modifications to cause misclassification into the target label from the source class than into other uninfected labels or other classes distinct from the source''}.

%\todo{Add more plots. Fix the captions. Maybe combining the same plot with multiple datasets using colors}\onote{yes, will be done.}

% \textit{ABS}~\cite{liu2019abs} improves Neural Cleanse by inspecting the neurons' activation change when providing different levels of stimulation. Since it performs similarly to Neural Cleanse, we expect similar results that we will address in future work.
% \onote{We need to have a convincing discussion about it. This is the weakest point of the section SP: I agree. We need better discussion}\gnote{I added ABS with NC, maybe now its better}

\subsection{Defense Methods Specific to the FL Setting}

FL specific methods, e.g., \textit{Krum}~\cite{blanchard2017machine}, \textit{FoolsGold}~\cite{fung2018mitigating}, \textit{Baffle}~\cite{andreina2021baffle}, \textit{CRLF}~\cite{xie2021CRLF}, and Fu et al.~\cite{fu2019attack} are adopted by the entire network and consider several assets of it. Krum is a robust aggregation mechanism that allows \textit{Byzantine clients} in the network; by using a majority-based scheme, aggregate the models which are closer, i.e., more similar. On the same basis, FoolsGold uses the cosine similarity between models to discard dissimilar ones. The intuition is that malicious clients collaborating to achieve the same goal have similar cosine models, which differ from honest clients. Regarding our threat model, assuming the server is malicious, defenses that require server participation do not hold in our settings, i.e., the FL network cannot assume that those will be implemented. Even more, the attack is not launched from the client, which is \textit{always} sharing an honest model. 
%\todo{what about the latest approaches? FLAME, FLGuard, etc.?}
Similarly, Baffle defends against backdoor attacks by a feedback-based voting mechanism, where every client tests their model with their dataset and submits the results to the server, which expects an incremental improvement in the accuracy. If most clients report a negative impact on the accuracy, the global model is flagged as malicious. This defense does not work on our attack setting since only one client is backdoored and would report that the accuracy has lowered. However, the majority of the clients would report high accuracy since they are not backdoored, causing a false negative. On the same basis, Fu et al.~\cite{fu2019attack} developed a robust aggregation algorithm using residual-based reweighting, which defends against backdoor attacks. This defense mechanism does not apply to our setup since the server supposed to aggregate with the new weighting system is malicious, and the clients cannot validate the correctness of the weighted aggregation.

\begin{figure}
    \centering
    \includegraphics[width=\columnwidth]{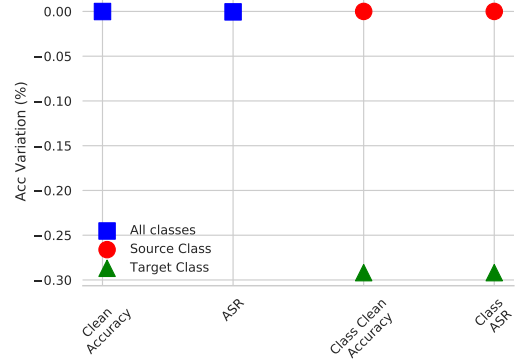}
    \caption{Accuracy variation after smoothing for $\epsilon=0.1$ MNIST $1\rightarrow7$ setting.}
    \label{fig:def_mnist_1_7}
\end{figure}

Xie et al.~\cite{xie2021CRLF} developed a clipping and smoothing method for defending against backdoor attacks in FL named CRLF. They theorized that norm clipping the model's weights at every epoch and adding Gaussian noise during training time in combination with randomized parameter smoothing at test time could prevent backdoors. We test this defense against our attack by cleanly training a CNN model for the MNIST and EMNIST datasets, keeping the same setting as in our approach, and for the defense, we set $\sigma=0.01$ to generate 1\,000 smoothed models and error tolerance of $\alpha=0.01$, the maximum norm to 100, and the clipping threshold to 15. After applying the smoothing, we observe no significant reduction, i.e., less than 1\%, on the ASR, see Figure~\ref{fig:def_mnist_1_7}. Our experiments show that smoothing by itself is not successful in preventing backdoor attacks. However, clipping cannot be applied since it requires an honest server, which cannot be guaranteed in our settings.

\subsection{Discussion}
\label{sec:discussion}

We tested our attack against the state-of-the-art defenses, and we now briefly discuss two potential evasion techniques against our attack and the limitations of our attack.

\begin{compactenum}
\item Backdoor models slightly degrade the accuracy of the model. In our approach, since just a subset of clients is backdoored, comparing the accuracy of every model will create inconsistency. Defenses following that approach from~\cite{wang2020model} assume that the server is trusted, which does not hold for our attack. However, adapting the defense to be executed by a trusted third party could be a feasible defense mechanism.

\item Differential privacy is widely used to prevent inference attacks~\cite{wei2020federated}. By inserting noise in the model's weights, model inversion attacks reconstruct noisy data. Since our attack relies heavily on reconstructing clients' datasets, this defense could make it difficult to launch our attack.
\end{compactenum}

Lastly, our attack is subject to some limitations. First, the attack is costly both in time and computational resources. Therefore, an attacker who lacks these may not be able to perform the attack. Second, despite many attacks considering the server as malicious, getting control of it could be more difficult than just assuming a malicious client. Therefore, our attack could not be implemented in scenarios where the server's trust is guaranteed.

\section{Conclusions \& Future Work}
\label{sec:conclusions}

This research studies the viability of client-wise targeted backdoor attacks and demonstrates the high performance of the attack. The first phase of our proposal investigated the creation of client data through a model inversion attack. The second part combined similarity matching and pixel pattern backdoor techniques to target a single client in an FL network. Our findings suggest that the combination of model inversion attacks and backdoors is a powerful duple, laying the groundwork for new threats. It also demonstrates that the client-targeted backdoor attack poses a real threat to an FL system, highlighting the importance of doing more exhaustive research and proposing specific defense strategies against them. Similarly, state-of-the-art defenses either do not succeed in defending our attack or do not hold for the setup. We, therefore, find reasonable the development of defense mechanisms that consider client-specific backdoors.

The generalizability of our results is subject to certain limitations. For instance, broader experimentation with different, more complex models, datasets, and numbers of clients is a natural progression of this work. Nevertheless, this study reinforces the idea that an attacker can cause severe degradation of the model in a targeted manner with little information.
Furthermore, we unravel two research directions to be addressed in the future. The first one is simplifying the attack's time and computational complexity, and making it suitable for a broader range of setups. The second one is relaxing the defined assumptions, thus empowering the attack and adapting it to more realistic scenarios. These findings could lead the research towards performing a client-wise targeted backdoor where a client is an attacker.

% This research studies the viability of client-wise targeted backdoor attacks and demonstrates the attack's high performance. The first phase of our proposal investigated the creation of client data through a model inversion attack. The second part combined different techniques such as similarity matching and pixel-pattern backdoor to target a single client in an FL network. Our findings suggest that the combination of model inversion attacks and backdoors is a powerful duple, laying the groundwork for new, more realistic attacks. It also demonstrates that the client-targeted backdoor attack poses a real threat to an FL system, highlighting the importance of doing more exhaustive research and proposing defense strategies against them. Overall, this study reinforces the idea that an attacker can cause severe degradation of the model in a targeted manner with little information. The findings are relevant for future work that further simplifies the considered assumptions, such as performing a client-wise targeted backdoor from a client's perspective as an attacker.

% Furthermore, future work that simplifies the attack in both time and computational complexity would make it suitable for a broader range of setups. The generalizability of these results is subject to certain limitations. For instance, broader experimentation with different, more complex models, datasets, and a broader number of clients is a natural progression of this work.

\bibliography{bibliography.bib}
\bibliographystyle{splncs04}

\appendices

\section{Reviews from Prior Submissions}

We have carefully gone over the reviewer's comments from prior submission (with the title "One is Enough: Client-Wise Targeted Backdoor in Federated Learning").
We found the comments very constructive to improve the quality of the paper. The paper has been revised thoroughly reflecting the changes and improvements we made based on the comments.
Here, we present the reviews and explain our improvements on each one separately.
	
{\underline{\bf Review 1}} 

The paper proposes a backdoor attack on the federated learning network that requires the compromise of a single client.  The idea is to leverage a Generative Adversarial Network to perform a model inversion attack, and then use a shadow federated learning aggregator to introduce the backdoor.

The idea is interesting, and the paper works on an important problem.

The writing of the paper can use some significant improvement, especially in Section 4.  The way it's written now seems like a laundry list with steps 1-2-3-4.  More intuitions could be added to explain why the steps are designed in such a way, and why this is the best way.  I also find that technical details are lacking in Section 4 in each individual component.

The contribution of the paper is more on an assemble of existing techniques, including GAN, SSN, and etc.  I find the idea of combining these techniques interesting, although each individual component is not new.

{\underline{\bf  Authors' Improvements 1}}

We rewrote Section IV by taking into account the feedback given by the reviewer.
In the current version, we have an overview section (Section IV.A) where we explain the techniques that are used and their reasoning.
Also, we added the necessary details in this section and also in the background section.
Finally, we clarified our contributions and emphasize the novelty in our attack regarding the scenario and the techniques that we utilized in Section I.A.

% 	\begin{center} 
% 		\fbox{
% 			\begin{minipage}{\columnwidth} 
% 				the text from the paper
% 			\end{minipage}
% 		}
% 	\end{center}

{\underline{\bf Review 2}} 

Summary of the paper:

This paper proposes a client-identified backdoor attack to federated learning,
where it identifies the client-target by GAN-based inference attack and Siamese
Neural Network training and then performs a client-specific backdoor attack. It
presents the detailed design of their method and conducts experiments to
evaluate the effectiveness of their backdoor attack.

Strengths:

The client-identified backdoor attack against federated learning models is an
interesting and important topic whose threat model is more practical than
previous works.

Weaknesses

The core of this paper is the client-identification process. However, there is
no verification of the identification effectiveness. I think the accuracy of
client identification should be added to make the proposed algorithms more
convincing.

Related works of backdoor attacks against federated learning are missing. I
think works should be included in Section 1.1. For example, [1-3]. It will be
more natural to introduce the motivation after discussing the limitations of
these works.

The evaluation lacks comparisons with related methods. It will be better to
provide results of baseline methods. For example, [1-3]. Besides, although it
discusses accuracy validation and differential privacy as defense methods, I
think more defense countermeasures are still necessary. Since it is a backdoor
attack, it should consider existing defenses against FL backdoors. For
example, [4-5]. It will be better to provide results of the attack
effectiveness after these defense methods have been adopted.

[1] Xie C, Huang K, Chen P Y, et al. Dba: Distributed backdoor attacks against
federated learning. International Conference on Learning Representations.
2019.

[2] Wang H, Sreenivasan K, Rajput S, et al. Attack of the tails: Yes, you
really can backdoor federated learning. Advances in Neural Information
Processing Systems, 2020, 33: 16070-16084.

[3] Sun Z, Kairouz P, Suresh A T, et al. Can you really backdoor federated
learning?. arXiv preprint arXiv:1911.07963, 2019.

[4] Xie C, Chen M, Chen P Y, et al. Crfl: Certifiably robust federated
learning against backdoor attacks. International Conference on Machine
Learning. PMLR, 2021: 11372-11382.

[5] Fu S, Xie C, Li B, et al. Attack-resistant federated learning with
residual-based reweighting. arXiv preprint arXiv:1912.11464, 2019.

{\underline{\bf  Authors' Improvements 2}}

Indeed, client identification is important to target only the victim, not all clients.
In the current version, we added the accuracy of the identification protocol for the non-IID dataset (that was used in the previous submission). In addition, we evaluated our attack (including the identification) with IID datasets as well. 
In both cases, our attack is successfully launched. 
See Section V for the experimental results.

We significantly improved the related work section (Section I.B) by including the aforementioned references [1-3] (corresponding to references [46,37,33] in the manuscript), and many other works.
Compared to the prior version, we have 14 new references.
Moreover, we compared our results with the baseline references given by the reviewer. The comparison results are explained in Section V.F.

We added a new section for the defenses (Section VI). 
In that section, we elaborately discuss the mentioned defense mechanisms and their effectiveness and suitability against our model. 
Specifically, in our model, since the server is malicious, the defense mechanisms utilizing honest aggregation (clipping and addition of noise in [4], and residual-based weighted aggregation in [5]) do not apply to our protocol. 
Nonetheless, we implemented the smoothing mechanism given in [4] and show that our attack is successful against the defense mechanism. A detailed explanation is given in Section VI. 

{\underline{\bf Review 3}} 

The paper proposes a technique to allow backdoor attacks to be client-targeted, compromising a single client while the rest remain unaltered. The attack uses a Generative Adversarial Network to perform a model inversion attack. Afterward, the attacker shadow-train the FL network by using a Siamese Neural Network, and backdoor the victim’s model.

Strength:

The topic is of importance and beneficial to the practice

Weakness:

•       Lack of sufficient novelty and technical contributions.

•       Weak experiments.

Detailed comments:

The proposed method was more or less a streamline implementation of off-the-shelf methods, resulting in limited technical contribution and novelty. In addition, I also wonder if it is necessary to inverse the model to synthesis the client’s training data. Since local model poisoning attack is a common strategy in which the adversary manipulates the local updates and has demonstrated strong attack capacity by a lot of literature, it would be possible to work on the clients’ gradient directly without model inversion.

Also, another major concern is about the experiments in which some essential settings are missing, and some of the results are incomplete. In regard to the experimental setting, the proposed strategy should work well in the non-IID setting (having said that, it seems that it only will work in non-iid setting, which is another weakness to me), while the experiment on non-IID setting is missing. In terms of the results presentation, the paper only outlines the results on MNIST, while the precise results on the other two datasets are missing.

{\underline{\bf  Authors' Improvements 3}}

We improved the contribution section to clarify the contributions and novelty of our work.
Unlike previous backdoor attacks in the FL setting, our attack targets a single client.
Therefore, we made several modifications to the existing off-the-shelf methods.
For example, we propose the first use of the SNN to identify the clients.
More specifically, we compare incoming anonymous updates and distinguish the victim client.
In the current version, we explain the details in Section IV.
Moreover, we use the \textit{model inversion} in the combination with a GAN for client identification. 

%Since we have a single target attack, we need to identify the targeted client from the anonymous model updates.

% In the current version of the paper, this is clarified.

We rewrote the experiments section (Section V).
We addressed the comments of the reviewer by clarifying the experimental settings and the precise results for all datasets. 
Moreover, in addition to existing results in non-IID settings, we did experiments with the IID case as well.
Our experimental results presented in Figures 5 and 6 show that our attack is successful in both non-IID and IID settings.

{\underline{\bf Review 4}} 

\# Paper summary

This paper proposes a new  backdoor attack against federated learning of Deep Neural Networks. More specifically, it identifies the client -target by a GAN and SNN. THen perform a client specific backdoor attack.

\# Strength

· The topic and the problem are interesting.

· clinet-wise backdoor attack attack on federated learning is severe and practical threaten.

\# Weakness

· Contribution is vague

· Lack of verifications

· The evaluation under state-of-the-art backdoor defenses is missing.

\# Detailed comments

In the Contribution section, the author gives many technical details about how they achieved their goal, but has no illustration of what's their contributions.  Also, there is no verification about the effectiveness of verification, in either theoretically or experimentally. As a Backdoor attack paper, the author didn't provide the comparison result to show the stealthiness of the proposed method against the SOTA backdoor defense[1] [2].

[1]Wang, Bolun, et al. "Neural cleanse: Identifying and mitigating backdoor attacks in neural networks." 2019 IEEE Symposium on Security and Privacy (SP). IEEE, 2019.

[2]Liu, Yingqi, et al. "Abs: Scanning neural networks for back-doors by artificial brain stimulation." Proceedings of the 2019 ACM SIGSAC Conference on Computer and Communications Security. 2019.

{\underline{\bf  Authors' Improvements 4}}

As we already stated previously, our contribution section (Section I.A) is significantly improved.
We focused on our contribution and the novelty rather than the technical details.  

To show the effectiveness of our attack, first, we added new experimental results (IID setting) and also implemented state-of-the-art defense mechanisms.
Secondly, in Section VI, we discussed several defense mechanisms and their applicability to our model.
Neural Cleanse [1] and ABS [2] defend against backdoor attacks by finding the small perturbations that make the model behave abnormally. 
Since both defense mechanisms work similarly,  we implemented the Neural Cleanse defense mechanism and showed that our attack is successful against it. 
A detailed discussion on the attacks and their effectiveness can be found in Section VI.

\end{document}